\documentclass[a4paper,oldversion]{aa} 
\usepackage{graphicx}
\usepackage{txfonts}
\begin{document}

   \title{The spatial clustering of X-ray selected AGN\\ in the
XMM-COSMOS field}

\author{
R. Gilli\inst{1}, 
G. Zamorani\inst{1},
T. Miyaji\inst{2},
J. Silverman\inst{3},
M. Brusa\inst{4},
V. Mainieri\inst{5},
N. Cappelluti\inst{4},
E. Daddi\inst{6},
C.~Porciani\inst{3},
L.~Pozzetti\inst{1},
F. Civano\inst{7},
A. Comastri\inst{1},
A. Finoguenov\inst{4},
F. Fiore\inst{8},
M. Salvato\inst{9},
C. Vignali\inst{10},
G. Hasinger\inst{4}, 
S. Lilly\inst{3},
C. Impey\inst{11},
J. Trump\inst{11},
P. Capak\inst{9},
H. McCracken\inst{12},
N. Scoville\inst{9},
Y. Taniguchi\inst{13},
C.~M.~Carollo\inst{3},
T.~Contini\inst{14},
J.-P.~Kneib\inst{15},
O. Le~Fevre\inst{15},
A.~Renzini\inst{16},
M. Scodeggio\inst{17},
S.~Bardelli\inst{1},
M.~Bolzonella\inst{1},
A.~Bongiorno\inst{4},
K.~Caputi\inst{3},
A.~Cimatti\inst{10},
G.~Coppa\inst{10},
O.~Cucciati\inst{18},
S.~de~la~Torre\inst{15},
L.~de~Ravel\inst{15},
P.~Franzetti\inst{17},
B.~Garilli\inst{17},
A.~Iovino\inst{18},
P.~Kampczyk\inst{3},
C.~Knobel\inst{3},
K.~Kova\v{c}\inst{3},
F.~Lamareille\inst{14},
J.-F.~Le~Borgne\inst{14},
V.~Le~Brun\inst{15},
C.~Maier\inst{3},
M.~Mignoli\inst{1},
R.~Pell\`o\inst{14},
Y.~Peng\inst{3},
E.~Perez~Montero\inst{14},
E.~Ricciardelli\inst{16},
M.~Tanaka\inst{5},
L.~Tasca\inst{15},
L.~Tresse\inst{15},
D.~Vergani\inst{1},
E.~Zucca\inst{1},
U.~Abbas\inst{15},
D.~Bottini\inst{17},
A.~Cappi\inst{1},
P.~Cassata\inst{15},
M.~Fumana\inst{17},
L.~Guzzo\inst{18},
A.~Leauthaud\inst{19},
D.~Maccagni\inst{17},
C.~Marinoni\inst{20},
P.~Memeo\inst{17},
B.~Meneux\inst{4},
P.~Oesch\inst{3},
R.~Scaramella\inst{8},
J.~Walcher\inst{12}
}

\authorrunning{R. Gilli et al.}
\titlerunning{The spatial clustering of AGN in the XMM-COSMOS field}

   \offprints{R. Gilli \\email:{\tt roberto.gilli@oabo.inaf.it}}

   \date{Received ... ; accepted ...}

\institute{ 
INAF -- Osservatorio Astronomico di Bologna, Via Ranzani 1, 40127
Bologna, Italy 
\and 
Instituto de Astronom\'ia, Universidad Nacional Aut\'onoma de
 M\'exico, Ensenada, M\'exico (mailing address: PO Box 439027, San
 Ysidro, CA, 92143-9027, USA)
\and 
Institute of Astronomy, Swiss Federal Institute of Technology (ETH
H\"onggerberg), CH-8093, Z\"urich, Switzerland
\and 
Max-Planck-Institut f\"ur extraterrestrische Physik, Postfach 1312,
D-85741 Garching, Germany
\and 
European Southern Observatory, Karl-Schwarzschild-Strasse 2, Garching,
D-85748, Germany
\and 
Laboratoire AIM, CEA/DSM - CNRS -
Universit\'e Paris Diderot, DAPNIA/SAp, Orme des Merisiers, 91191
Gif-sur-Yvette, France 
\and 
Harvard-Smithsonian Center for
Astrophysics, 60 Garden Street, Cambridge, MA 02138, USA 
\and 
INAF -- Osservatorio Astronomico di Roma, Via di Frascati 33, 
00040 Monte Porzio Catone (Roma), Italy 
\and 
California Institute of Technology, MC 105-24, 1200 East California
Boulevard, Pasadena, CA 91125, USA
\and 
Dipartimento di Astronomia, Universit\`a degli Studi di Bologna, 
Via Ranzani 1, 40127 Bologna, Italy 
\and 
Steward Observatory, University of Arizona, Tucson, AZ 85721, USA
\and 
Institut d'Astrophysique de Paris, 98 bis Boulevard Arago, F-75014
Paris, France
\and 
Research Center for Space and Cosmic Evolution, Ehime
University, Bunkyo-cho 2-5, Matsuyama 790-8577, Japan 
\and 
Laboratoire d'Astrophysique de Toulouse-Tarbes, Universit\'e de
Toulouse, CNRS, 14 avenue E. Belin, F-31400 Toulouse, France
\and 
Laboratoire d'Astrophysique de Marseille, CNRS-Universit\'e
de Provence, Traverse du Siphon, BP 8, 13012 Marseille, France
\and 
Dipartimento di Astronomia, Universit\`a di
Padova, Vicolo Osservatorio 2, 35122 Padova, Italy
\and 
INAF -- Istituto di Astrofisica
Spaziale e Fisica Cosmica, Via Bassini 15, I-20133, Milan, Italy
\and 
INAF -- Osservatorio Astronomico di Brera, I-23807, Merate (LC), Italy 
\and 
University of California, Lawrence Berkeley National Laboratory, 
1 Cyclotron Rd, Berkeley, CA 94720, USA
\and 
Centre de Physique Th\'eorique, UMR 6207 CNRS-Universit\'e de
Provence, Case 907, 13288 Marseille, France}

\abstract{

We study the spatial clustering of 538 X-ray selected AGN in the 2
deg$^2$ XMM-COSMOS field that are spectroscopically identified with
$I_{AB}<23$ and span the redshift range $z=0.2-3.0$. The median
redshift and luminosity of the sample are $z = 0.98$ and
$L_{0.5-10}=6.3\times 10^{43}$ erg s$^{-1}$, respectively. A strong
clustering signal is detected at $\sim 18\sigma$ level, which is the
most significant measurement obtained to date for clustering of X-ray
selected AGN. By fitting the projected correlation function $w(r_p)$
with a power law on scales of $r_p=0.3-40\,h^{-1}$ Mpc, we derive a
best-fit comoving correlation length of $r_0 = 8.6\pm0.5\,h^{-1}$ Mpc
and slope of $\gamma=1.88\pm0.07$ (Poissonian errors; bootstrap errors
are about a factor of 2 larger). An excess signal is observed in the
range $r_p\sim5-15\,h^{-1}$ Mpc, which is due to a large-scale
structure at $z\sim 0.36$ containing about 40 AGN, a feature which is
evident over many wavelengths in the COSMOS field. When removing the
$z\sim 0.36$ structure or computing $w(r_p)$ in a narrower range
around the peak of the redshift distribution (e.g. $z=0.4-1.6$), the
correlation length decreases to $r_0 \sim 5-6\,h^{-1}$ Mpc, which is
consistent with what is observed for bright optical QSOs at the same
redshift.

We investigate the clustering properties of obscured and unobscured
AGN separately, adopting different definitions for the source
obscuration. For the first time, we are able to provide a significant
measurement for the spatial clustering of obscured AGN at $z\sim 1$.
Within the statistical uncertainties, we do not find evidence that AGN
with broad optical lines (BLAGN) cluster differently from AGN without
broad optical lines (non-BLAGN).
Based on these results, which are limited by object statistics,
however, obscured and unobscured AGN are consistent with inhabiting
similar environments.

The evolution of AGN clustering with redshift is also investigated. No
significant difference is found between the clustering properties of
XMM-COSMOS AGN at redshifts below or above $z=1$.

The correlation length measured for XMM-COSMOS AGN at $z\sim 1$ is
similar to that of massive galaxies (stellar mass $M_\star\gtrsim
3\times 10^{10} \; M_{\odot}$) at the same redshift. This suggests
that AGN at $z\sim 1$ are preferentially hosted by massive galaxies,
as observed both in the local and in the distant ($z\sim 2$)
Universe. According to a simple clustering evolution scenario, we find
that the relics of AGN are expected to have a correlation length as
large as $r_0 \sim 8\,h^{-1}$ Mpc by $z=0$, and hence to be hosted by
local bright ($L\sim L_\star$) ellipticals.

We make use of dark matter halo catalogs from the Millennium
simulation to determine the typical halo hosting moderately luminous
$z\sim 1$ AGN. We find that XMM-COSMOS AGN live in halos with masses
$M\gtrsim 2.5\times 10^{12}\;M_{\odot}\; h^{-1}$. By combining the
number density of XMM-COSMOS AGN to that of the hosting dark matter
halos we estimate the AGN duty cycle and lifetimes. We find lifetimes
approximately of 1 Gyr for AGN at $z\sim 1$, which are longer than
those estimated for optically bright QSOs at the same redshift. These
longer lifetimes mainly reflect the higher number density of AGN
selected by X-ray samples.

   \keywords{Surveys -- Galaxies: active -- X-rays: general --
		Cosmology: large-scale structure of Universe} }

   \maketitle

\section{Introduction} \label{introduction}

Several pieces of evidence point towards an intimate correlation
between the evolution of galaxies and the accretion and growth of
supermassive black holes (SMBHs) at their centers, indicating that
most galaxies in the Universe spent a fraction of their lifetimes as
active galactic nuclei (AGN). In the local Universe, most galaxy
bulges indeed host a supermassive black hole (see e.g. Ferrarese \&
Ford, \cite{ff05} for a review), whose mass scales with the bulge mass
and stellar velocity dispersion (Ferrarese \& Merritt 2000, Gebhart et
al. 2000). Furthermore, the growth of SMBHs during active accretion
phases, which is traced by the cosmological evolution of the AGN
luminosity function (Ueda et al. 2003, Hasinger et al. 2005, La Franca
et al. 2005, Silverman et al. 2008a), has been shown to eventually
match the mass function of SMBHs in the local Universe (e.g. Marconi et
al. \cite{marconi}; Yu \& Tremaine \cite{yu}, Shankar et al. 2004).

While the SMBH vs galaxy co-evolution is now an accepted scenario, the
details of this joint evolution are not fully understood yet. Nuclear
activity in bright QSOs is thought to be induced by major mergers or
close encounters of gas-rich galaxies in the context of hierarchical
structure formation (e.g. Kauffmann \& Haehnelt 2000, Cavaliere \&
Vittorini 2002, Hopkins et al. 2006). Alternatively, nuclear activity
can be simply related to the physical processes (e.g. star formation)
going on in a single galaxy, without being induced by mergers or
interactions with neighboring objects (e.g. Granato et al. 2004).
Overall, the role played by the environment in triggering both nuclear
activity and star formation is still a matter of debate.

Just like local ultraluminous infrared galaxies (Sanders \& Mirabel
1996), the population of bright submillimeter sources at $z\sim 2$
(Chapman et al. 2003) is hosting both star formation and nuclear
activity (Alexander et al. 2005), as a result of galaxy interactions
(Tacconi et al. 2008). 
However, the majority of $z\sim 2$ AGN selected at faint X-ray fluxes
seem to be hosted by galaxies with a spectral energy distribution
typical of passively evolving objects (Mainieri et al. 2005). The
concurrent growth of black holes and stellar mass has been observed in
IR galaxies at $z\sim 2$ by Daddi et al. (2007), who suggested a
long-lived ($>0.2$ Gyr) AGN plus starburst phenomenon, unlikely to be
triggered by rapid merger events. In the local galaxies observed by
the Sloan Digital Sky Survey (SDSS, York et al. 2000), nuclear activity
does not appear to be correlated to the presence of close companions,
while star formation does (Li et al. 2008). A common merger origin for
both phenomena cannot be ruled out, however, provided they occur at
different times (see Li et al. 2008).


The relation between nuclear activity and the environment can be
studied via clustering techniques in the context of large-scale
structure formation, in which the growth of baryonic structures is
supposed to follow the formations of dark matter halos (DMHs).


The comparison between the clustering properties of AGN and those of
DMHs predicted by cold dark matter (CDM) models can be used to
evaluate the typical mass of the DMHs in which AGN form and reside as
a function of cosmic time. The most recent measurements have shown
that bright QSOs in the redshift range 0-3 reside into DMHs of mass
$M>10^{12} \: M_{\odot}$ (Grazian et al. \cite{grazi04}; Porciani et
al. 2004, Croom et al. \cite{croom05}; but see Padmanabhan et al. 2008
for lower mass estimates at $z\sim0.3$).
In addition, the ratio between the AGN space density and the
space density of host DMHs may provide an estimate of the AGN lifetime
(e.g. Martini \& Weinberg \cite{mw01}). Current estimates are largely
uncertain, constraining the AGN lifetime in the range of a few $\times
10^6 - 10^8$ yr (Grazian et al. \cite{grazi04}; Porciani et
al. \cite{porc04}). Finally, the comparison between the clustering
properties of different galaxy types and AGN can be used to estimate
AGN hosts and to estimate the descendant and progenitors of
AGN at any given redshifts.

AGN clustering has been traditionally studied by means of the
two-point correlation function applied to optically selected QSO
samples (e.g. Shanks et al. \cite{shank87}, La Franca et
al. \cite{lafra98}). The most recent and solid results of these
analyses come from the two largest QSO surveys to date, namely the 2dF
QSO Redshift Survey (2QZ, e.g. Croom et al. \cite{croom05}), and the
SDSS. The 2QZ is based on a sample of more than 20000 objects with
redshifts $0.2\lesssim z \lesssim 3.0$. When calculating the
correlation function in real space and approximating it with a
power law $\xi(r)=(r/r_0)^{-\gamma}$, the QSO correlation length and
slope were found to be $r_0=5.0 \pm 0.5 \:h^{-1}$ Mpc and
$\gamma=1.85\pm 0.13$ at a median redshift of $\bar z=1.5$ (Da Angela
et al. \cite{daa05}). Some evidence of a flattening towards smaller
scales was also reported, with $\gamma=1.45$ at projected scales below
$10\: h^{-1}$ Mpc (Da Angela et al. \cite{daa05}). The clustering
level of 2QZ QSOs is similar to that of early type galaxies at the
same redshift (Coil et al. 2004, Meneux et al. 2006), suggesting they
reside in environments of similar density. A tentative detection of
$z\sim 1$ AGN residing preferentially in the same environment of blue
rather than red galaxies has been reported by Coil et al. (2007). The
QSO clustering is observed to be a strong function of redshift (Croom
et al. 2005, Porciani \& Norberg 2006), with the correlation length of
luminous QSOs at $z\sim 4$ being as high as $r_0=24\,h^{-1}$ Mpc (Shen
et al. 2007). This suggests that luminous, early QSOs are hosted by
the most massive and rare DMHs and hence form in the highest density
peaks of the dark matter distribution. The evidence of luminosity
dependent clustering is, on the contrary, still marginal (Porciani \&
Norberg 2006).

The above results are mostly based on AGN selected by means of their
blue optical colors and broad optical lines; i.e., they essentially
refer to unobscured, type 1 AGN. With the notable exception of the
measurement performed by the SDSS on a local sample of narrow-line AGN
(Li et al. 2006), to date there has been no information on the clustering
properties of obscured AGN, which, based on the results from deep
X-ray surveys (Brandt \& Hasinger 2005, Tozzi et al. 2006) and X-ray
background synthesis models (e.g. Gilli, Comastri \& Hasinger 2007a),
are found to be a factor of $\gtrsim 4$ more numerous than unobscured
ones; i.e., they are the most abundant AGN population in the Universe,
dominating the history of accretion onto SMBHs (e.g. Fabian 1999). If
the unified model strictly applies, i.e. the nuclear obscuration is
just an orientation effect, one should not expect differences in the
clustering properties of obscured and unobscured AGN. However, several
exceptions to the strict unified model are known. Source obscuration
is in many cases related to the gas content and evolutionary stage of
the host galaxy, rather than to a small-scale torus intercepting the
line of sight (Malkan et al. 1998). Models have been proposed in which
the onset of nuclear activity starts embedded in an envelope of gas
and dust, which is later on swept out by the QSO radiation (see
e.g. Hopkins et al. 2006). If this were the case, obscured and
unobscured AGN would be just two subsequent stages along a galaxy
lifetime. The different durations of these two stages and their
relation with the environment may produce different clustering
properties between obscured and unobscured AGN.

One obvious way to obtain samples of obscured AGN is through X-ray
observations. Besides reducing the obscuration bias dramatically,
especially in the hard 2-10 keV band, X-ray selection also has the
advantage of being effective in selecting distant low-luminosity AGN,
whose optical light is diluted by the host galaxy emission and
therefore missed by color-based optical surveys.


In the past years the limited sample size of X-ray selected AGN
prevented clustering analyses as detailed as for optically selected
objects. In particular, the lack of dedicated optical follow-up
programs of X-ray sources providing large samples with spectroscopic
measurements, has not allowed accurate estimates of the spatial
clustering of X-ray selected AGN, limiting most studies to angular
clustering. Numerous investigations of the two point angular
correlation function of X-ray sources have indeed appeared in the
literature, but the results suffer from rather large
uncertainties. Early attempts to measure the angular clustering of
X-ray selected sources were performed by Vikhlinin et al. (1995) and
Carrera et al. (1998) based on ROSAT pointings. More recent results
based on Chandra and XMM data have been obtained by Basilakos et
al. (2004), Gandhi et al. (2006), Puccetti et al. (2006), Miyaji et
al. (2007), Carrera et al. (2007), Plionis et al. (2008), and Ueda et
al. (2008). In particular, Miyaji et al (2007) and Gandhi et
al. (2006) have computed the angular correlation function over
contiguous areas of a few square degrees (the 2 deg$^2$
COSMOS field and the $\sim$4 deg$^2$ XMM-LSS field, respectively),
which should reduce the impact of cosmic variance. In the COSMOS field,
Miyaji et al. (2007) measured a correlation length of about $10\pm
1\;h^{-1}$ Mpc, while only a loose constraint ($r_0=3-9\;h^{-1}$) Mpc
was obtained in the XMM-LSS by Gandhi et al. (2006). Very recently an
attempt to measure the angular clustering of high-redshift ($z\sim 3$),
X-ray selected AGN has been done by Francke et al. (2008).

The few examples of spatial clustering of X-ray selected sources
appeared in the literature are limited by low statistics. Based on a
sample of $\sim 220$ QSOs at $z\sim 0.2$ found in the 80 deg$^2$ ROSAT
North Ecliptic Pole survey (NEP, Gioia et al. \cite{gioia03}), Mullis
et al. (\cite{mullis04}) were able to measure a correlation signal to
$\gtrsim 3\sigma$ level. By fixing the correlation slope to
$\gamma=1.8$, they found a best-fit correlation length of
$r_0=7.4^{+1.8}_{-1.9}\:h^{-1}$ Mpc. Because of the relatively short
exposures in the NEP survey and the limited ROSAT sensitivity and
bandpass (0.1-2.4 keV), only bright, luminous, unobscured QSOs have been
detected in this sample (median $L_{0.5-2{\rm keV}}\sim 9\times
10^{43}$ erg s$^{-1}$, corresponding to $L_{0.5-10{\rm keV}}\sim
2\times 10^{44}$ erg s$^{-1}$ for a spectral slope of
$\alpha=0.9$). Later, by analyzing data from the {\it Chandra} Deep
Field South (CDFS, Rosati et al. \cite{rosati02}) and Chandra Deep
Field North (CDFN, Alexander et al. \cite{alex03}), Gilli et
al. \cite{gilli05} were able to detect clustering at $>7\sigma$ level
for $z\sim 0.7$ AGN with $L_{0.5-10{\rm keV}}\sim 10^{43}$ erg s$^{-1}$.
However, the best-fit correlation length was found to vary by a factor
of $\sim 2$ between the two fields ($r_0=10\,h^{-1}$ Mpc in the CDFS,
$r_0=5\,h^{-1}$ Mpc in the CDFN), revealing strong cosmic variance
over these small, 0.1 deg$^2$ each, sky areas. Although with limited
significance ($\sim 3\sigma$), in the CDFs, it was also possible to
determine the clustering properties of obscured AGN only, which did
not show significant differences with respect to those of unobscured
ones within the uncertainties (Gilli et al. \cite{gilli05}). The most
recent measurement is the one performed in the larger, 0.4 deg$^2$
field covered by the CLASXS (Yang et al. 2006). A correlation length
of $\sim 5.7\,h^{-1}$ Mpc was found for X-ray selected AGN at $z\sim
1.2$, with average luminosity of $L_{0.5-10{\rm keV}}\sim 6\times
10^{43}$ erg s$^{-1}$.

A large number of X-ray surveys are ongoing and are expected to
provide larger samples of sources over wide sky areas and with
different limiting fluxes. They will allow studies of clustering of
AGN in different redshift and luminosity regimes. A few examples of
such surveys are X-Bootes (Murray et al. 2005), XMM-LSS (Pierre et
al. 2007), Extended CDFS (Lehmer et al 2005), AEGIS (Nandra et
al. 2005), and XMM-COSMOS (Hasinger et al. 2007). One of these
samples, the XMM survey in the 2 deg$^2$ COSMOS field (XMM-COSMOS),
has been specifically designed to study with the best statistics the
clustering of X-ray selected AGN. One of the main goals of XMM-COSMOS
was indeed to provide the best measurement to date of the correlation
function of X-ray selected AGN and to allow a reliable measurement of
the correlation function of obscured AGN at $z>0$ for the first time.

The optical spectroscopic identification of the 1822 pointlike X-ray
sources detected by XMM-COSMOS continues. In this paper we present the
results based on the first third of the objects. The paper is
organized as follows. In Sect. 2 we summarize the X-ray and optical
follow-up observations of the XMM-COSMOS sample and present the source
catalog used in our analysis. In Sect. 3 we describe the methods of
estimating the correlation function of X-ray selected sources. In
Sect. 4 several safety checks are performed to validate the adopted
techniques. The results of our analysis are presented in Sect. 5. In
Sect. 6 the results are discussed and interpreted. The conclusions and
prospects for future work are finally presented in Sect. 7.

Throughout this paper, a flat cosmology with $\Omega_m=0.25$ and
$\Omega_{\Lambda}=0.75$ is assumed (Spergel et
al. \cite{spergel}). For comparison with previous measurements we
refer to correlation lengths and distances in units of $h^{-1}$ Mpc
comoving, where $H_0=100\;h$ km s$^{-1}$ Mpc$^{-1}$. Masses of dark
matter halos are also expressed in units of $h^{-1} \; M_{\odot}$ for
consistency with the Millennium simulation, and AGN and halo space
densities are expressed in units of $h^3$ Mpc$^{-3}$. AGN luminosities
and lifetimes are calculated using $h=0.7$.

\begin{figure}[t]
\includegraphics[width=9cm]{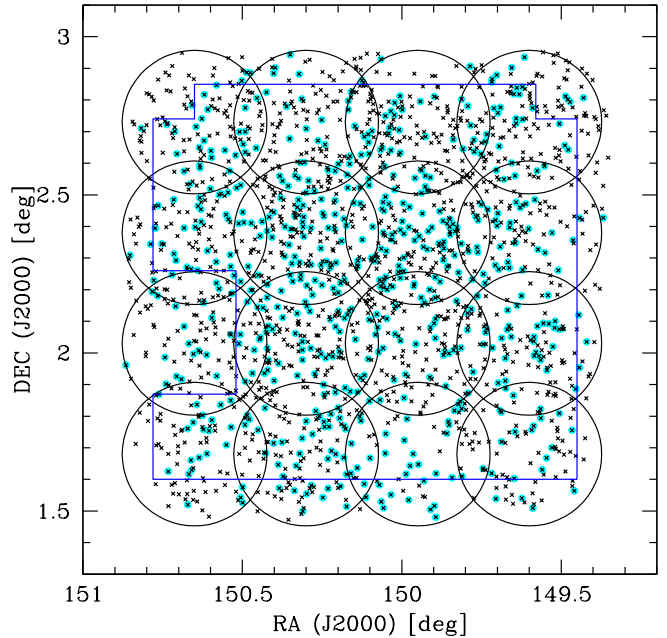}
\caption{Distribution on the sky of the 1822 pointlike sources detected by XMM
(crosses). Spectroscopically identified objects are shown as
filled circles. The area covered by spectroscopic observations is also
shown: the large circles show the 16 Magellan IMACS pointings, while the
polygon represents the area presently covered by zCOSMOS observations.}
\label{sky}
\end{figure}

\section{The sample} \label{xray}

The XMM-COSMOS survey is part of the COSMOS legacy project (see
Scoville et al. 2007a for an overview of the survey), an extensive
multiwavelength campaign to observe a $1.4\times 1.4$ deg equatorial
field centered at $(RA,DEC)_{J2000}=(150.1083, 2.210)$. A number of
large programs with the major observing facilities have been already
performed or are ongoing, including HST (Scoville et al. 2007b), VLT
(Lilly et al. 2007), SUBARU (Taniguchi et al. 2007), VLA (Schinnerer
et al. 2007), XMM (Hasinger et al. 2007), Chandra (Elvis et al. in
preparation) Spitzer (Sanders et al. 2007), and GALEX (Zamojski et
al. 2007; Schiminovich et al. in preparation).

XMM-COSMOS is a mosaic made of 53 partially overlapping XMM pointings
that cover the entire 2 deg$^2$ COSMOS field. The XMM observations were
allocated across two announcements of opportunities (AO-4 and AO-5)
and performed in two different passes\footnote{We do not consider here
two additional pointings performed separately in AO-6.}, for a total of
1.4 Ms exposure time. Each pass was arranged into a regular grid of
$\sim 30$ ks pointings separated by 8 arcmin each to cover the full 2
deg$^2$ field. In the second pass the grid pattern was shifted by 1
arcmin with respect to the first pass to ensure maximum uniformity in
the sensitivity over the final mosaic. The limiting fluxes reached in
the regions of maximum exposure are $5\times 10^{-16}$, $3\times
10^{-15}$, $6\times 10^{-15}$ erg cm$^{-2}$ s$^{-1}$ in the 0.5-2,
2-10, and 5-10 keV, respectively, while the entire 2 deg$^2$ area is
covered down to $2.4\times 10^{-15}$, $1.5\times 10^{-14}$, $2\times
10^{-14}$ erg cm$^{-2}$ s$^{-1}$ in the same bands. In total 1822
pointlike sources have been detected in at least one band down to a
likelihood threshold of 10 (see Cappelluti et al. 2007 for the source
detection process). The final catalog will be presented in a
forthcoming paper (Cappelluti et al. in preparation). A number of
results concerning the first pass (0.8 Msec total exposure) have been
published by Cappelluti et al. (2007), Miyaji et al. (2007), and Mainieri
et al. (2007).

The optical identification of XMM sources is currently in progress
(see Brusa et al. 2007 for the initial results based on the first 12
XMM pointings). Unique optical and/or infrared counterparts for most
($\sim$88\%) of the XMM sources have now been recognized. Thanks to
the Chandra observations in the central COSMOS square deg (Elvis et
al. in prep), a number of formerly ambiguous identifications have now
been made secure (Brusa et al. in preparation). The main dedicated
spectroscopic follow-up programs of XMM sources are being conducted
with the IMACS instrument at the 6m Magellan telescope (Trump et
al. 2007) and with VIMOS at the VLT within the zCOSMOS program
(e.g. Lilly et al. 2007). A number of spectroscopic redshifts were also
obtained by cross-correlating the XMM catalog with published
spectroscopic catalogs like the SDSS. About 46\% of the total
spectroscopic sample were obtained with only IMACS observations,
25\% with only zCOSMOS, and another 24\% has been observed in both
programs. The remaining 5\% were obtained by cross-correlation
with public catalogs. Quality flags were assigned to the
redshifts measured by IMACS and zCOSMOS. We considered here only the
621 X-ray pointlike sources with highest quality flags, which have been
identified as extragalactic objects. By considering the 150 duplicated
redshifts (i.e. those objects observed by both IMACS and VIMOS) the
accuracy in the redshift measurements is verified to be
$\sigma_z<0.002$.

Since we combine measurements obtained with different optical
instruments by different groups, which adopt somewhat different
spectral classification criteria, it is not easy to obtain a uniform
classification scheme for the optical counterparts of XMM sources. The
only class with a common definition in the various catalogs is
that of objects showing broad emission lines (FWHM$\gtrsim 2000$
km/s). Therefore, in the following, as far as optical classification
is concerned, we simply divide the sample into objects with and
without broad optical lines, referring to them as broad line AGN
(BLAGN) and non-BLAGN, respectively (see also Brusa et al. 2007).

\begin{figure}[t]
\includegraphics[width=9cm]{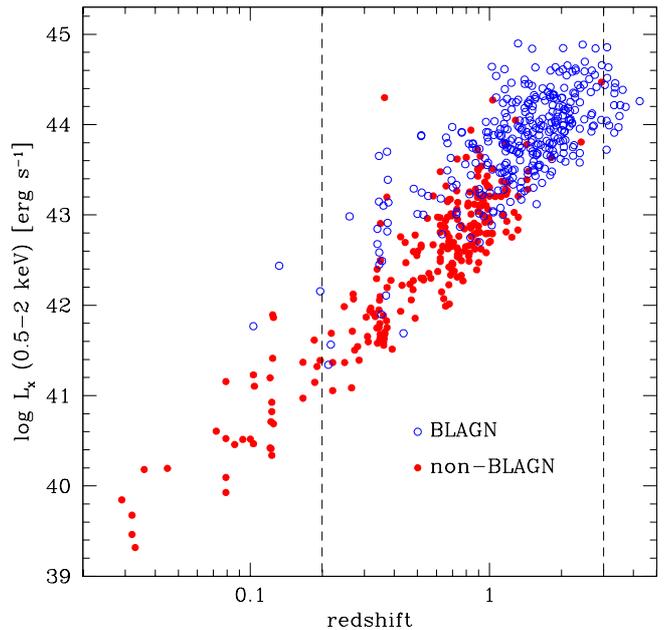}
\caption{Soft X-ray luminosity vs redshift relation for the
spectroscopically identified sources in XMM-COSMOS. Broad line AGN
(BLAGN) and non broad line AGN (non-BLAGN) are shown as blue open
circles and red filled circles, respectively. Only objects in the
redshift range $z=0.2-3$ (vertical dashed lines) have been considered
for the clustering analysis.}
\label{lxz}
\end{figure}

\begin{figure}[t]
\includegraphics[width=9cm]{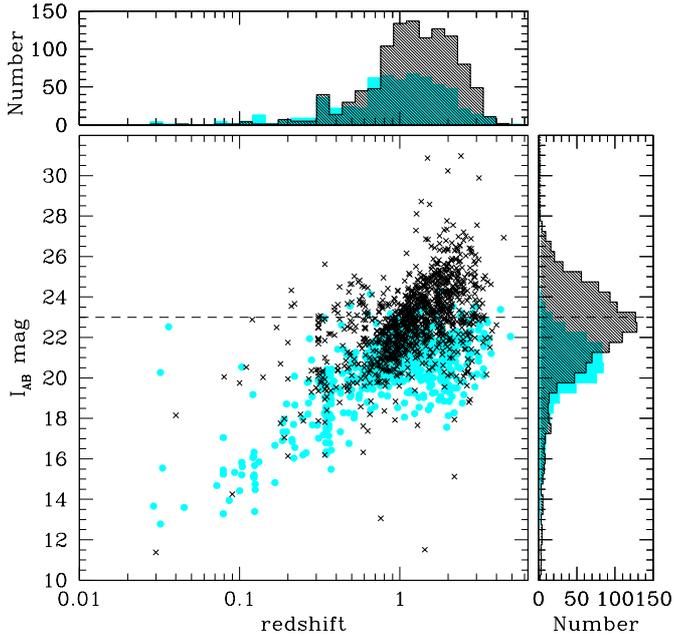}
\caption{I-band magnitude vs redshift distribution of the optical
counterparts of XMM sources. Objects with spectroscopic redshift
or only photometric redshift are shown as filled circles or
crosses, respectively. The dashed horizontal line shows the
$I_{AB}=23$ magnitude limit considered for the clustering analysis
(see text).}
\label{iz}
\end{figure}

The 0.5-2 keV X-ray luminosity vs redshift distribution of BLAGN and
non-BLAGN, is shown in Fig.~\ref{lxz}. BLAGN are on average observed
at higher redshift and at luminosities above $10^{42.5}$ erg s$^{-1}$,
while non-BLAGN are observed down to very low X-ray luminosities and
may therefore include a significant fraction of normal galaxies. To
exclude from the sample those objects that are not likely to be AGN, we
then considered only those sources at a redshift higher than 0.2. As
shown in Fig.~\ref{lxz}, this cut essentially removes most
low-luminosity objects, leaving in the sample only objects with
$L_{0.5-2}\gtrsim 10^{41.5}$ erg s$^{-1}$. \footnote{We verified that
adopting a luminosity cut at $L_{0.5-10}>10^{41.5}$ erg s$^{-1}$
produces similar results as adopting a redshift cut at z=0.2.} In
addition we considered for our analysis only objects at redshifts
below 3, since beyond this limit the source density becomes extremely
low and the selection function is very uncertain (see Sect. 3).

By exploiting the large multiwavelength database available in
COSMOS, Salvato et al. (2008) are able to estimate a photometric
redshift for $\sim 85\%$ of the XMM sources. For the remaining 15\% of
the sample, the main reason for not attempting a photometric redshift
estimate was the ambiguity in the correct association with the
optical/IR counterpart (see Brusa et al. 2007). This issue, however,
is not expected to introduce any bias related to source
distances. Therefore the redshift distribution estimated using
photometric redshifts should be very close to that of the entire XMM
sample. On the contrary, spectroscopic redshifts have been measured
for a minority ($\sim36\%$) of the total XMM sample (including 34
stars), making it possible that objects with measured
spectroscopic redshift are not a fair representation (i.e. a random
sampling) of the total AGN population detected by XMM. The I-band
magnitude vs redshift distribution of the objects with spectroscopic
redshifts and only photometric redshifts is shown in
Fig.~\ref{iz}. Objects only with photometric redshift are on average
optically fainter and at higher redshift than objects with
spectroscopic redshifts. A Kolmogorov-Smirnov test performed on the
redshifts distribution of objects with and without spectroscopic
redshift indicates that they differ at $>3\sigma$ level; i.e., the
spectroscopic sample is not a fair representation of the entire AGN
population detected by XMM. We therefore impose a magnitude cut at
$I_{AB}<23$, which excludes only a small fraction ($<3\%$) of
spectroscopically identified objects but increases the spectroscopic
completeness to about 60\%. The redshift distribution of objects
brighter than $I_{AB}=23$ with and without spectroscopic redshifts are
statistically indistinguishable. Therefore, we conclude that, for
objects with $I_{AB}<23$, the spectroscopic selection does not include
any bias against high-redshift objects.

\begin{figure}[t]
\includegraphics[width=9cm]{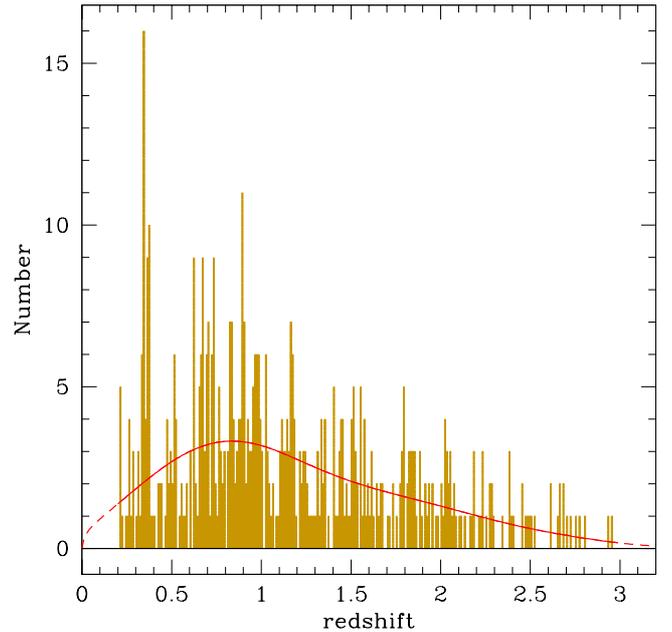}
\caption{Redshift distribution for the 538 X-ray sources in the
reference sample in bins of $\Delta z = 0.01$. The solid curve is
obtained by smoothing the observed redshift distribution with a
Gaussian with $\sigma_z=0.3$ and is used to generate the random
control sample in the correlation function estimate.}
\label{zdist}
\end{figure}

In the following we consider the sample of 538 XMM objects with
$I_{AB}<23$ and spectroscopic redshift in the range $z=0.2-3.0$ as our
reference sample. The average redshift and X-ray luminosity of this
sample are $z=0.98$ and $L_{0.5-10}=6.3\times 10^{43}$ erg/s,
respectively. The redshift distribution of the sample sources is shown
in Fig.~\ref{zdist}. A number of redshift structures are observed, the
most prominent of which is at $z\sim 0.36$, also observed at other
wavelengths in COSMOS (Lilly et al. 2007).
 


\begin{figure}[t]
\includegraphics[width=8.5cm]{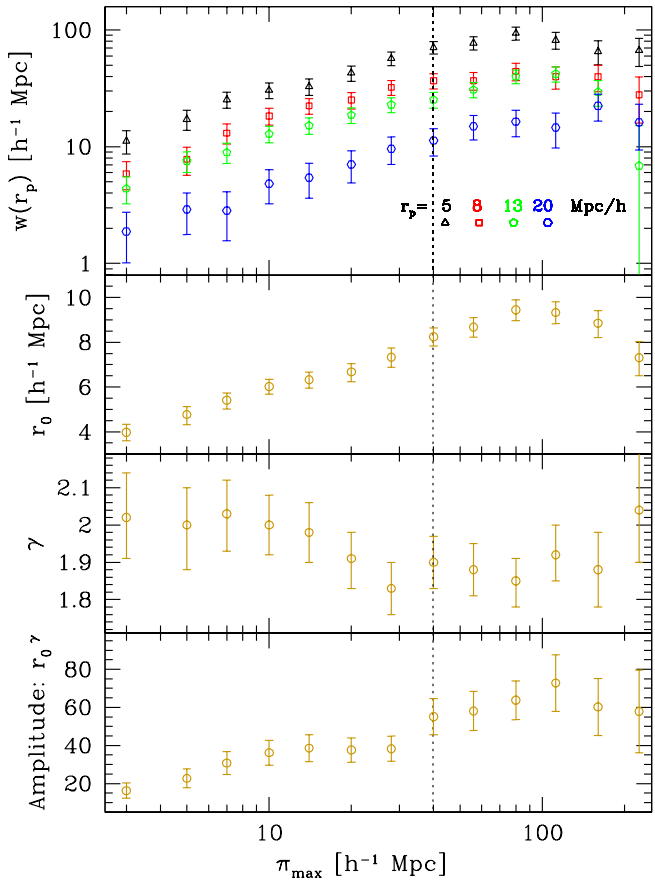}
\caption{{\it Upper panel:} projected correlation function $w(r_p)$
computed at different $r_p$ scales (see label) as a function of the
integration radius $\pi_{max}$. From {\it top middle} to {\it bottom
panels} the correlation length $r_0$, slope $\gamma$, and amplitude
$r_0^{\gamma}$ are shown as a function of $\pi_{max}$, respectively.}
\label{rov}
\end{figure}

\section {Analysis techniques}

The basic statistics commonly used to measure the clustering
properties of galaxies is the two point correlation function $\xi(r)$,
defined as the excess probability over random of finding a pair with
one object in the volume $dV_1$ and the second in the volume $dV_2$,
separated by a distance $r$ (Peebles \cite{peeb80}):

\begin{equation}
dP = n^2[1+\xi(r)]dV_1dV_2,
\end{equation}
where $n$ is the mean object space density. In our calculations we
always refer to comoving distances and volumes. In a flat
Universe one can simply estimate the comoving distance between two
objects at redshifts $z_1$ and $z_2$ separated on the sky by an angle
$\theta$ with the cosine rule (see e.g. Osmer et al. \cite{osmer81}):

\begin{equation}
s=\sqrt{{d_1}^2+{d_2}^2-2d_1d_2\rm{cos}\theta},
\end{equation}
where $d_i$ is the radial comoving distance to object $i$, which, again
in a flat universe, reads as

\begin{equation}
d_i = {c\over{H_0}}\int_0^{\,z_i}dz\,[\Omega_m(1+z)^3+\Omega_{\Lambda}]^{-1/2}.
\end{equation}

However, a well known feature of Eq. 2 in measuring pair separations
is that it is affected by redshift space distortions since peculiar
velocities combine with the source distance to produce the measured
redshift. (This is the same effect which in redshift surveys transforms
galaxy clusters into fingers-of-God). The net effect of peculiar
velocities is to shift pairs from smaller to larger radial
separations, thus shifting the clustering power towards higher
scales. In particular a flatter slope with respect to the real space
correlation function is generally observed.
A similar effect is also produced by the uncertainties in the redshift
measurements.

To overcome these problems one can resort to the so-called projected
correlation function:

\begin{equation}
w(r_p) = \int_{-\pi_{max}}^{\pi_{max}} \xi(r_p, \pi)d\pi,
\label{wdef}
\end{equation}
where $\xi(r_p, \pi)$ is the two-point correlation function expressed
in terms of the separations perpendicular ($r_p$) and parallel ($\pi$)
to the line of sight as defined in Davis \& Peebles (\cite{dp83}) and
applied to comoving coordinates:

\begin{equation}
\pi = |d_1 - d_2|\:,\: r_p=(d_1+d_2)\rm{tan}{\theta\over2}\:.
\end{equation}

Since $w(r_p)$ is an integral along the radial coordinate, it is
independent of peculiar velocity effects and can therefore be used as
an estimate of the true, real-space correlation function. In
particular, if the real space correlation function can be approximated
by a power law of the form $\xi(r)=(r/r_0)^{-\gamma}$ and $\pi_{max}=
\infty$, then the following relation holds (Peebles \cite{peeb80}):

\begin{equation}
w(r_p) =A(\gamma) r_0^{\gamma} r_p^{1-\gamma},
\label{wpow}
\end{equation}
where $A(\gamma)=\Gamma(1/2)\Gamma[(\gamma-1)/2]/\Gamma(\gamma/2)$
and $\Gamma(x)$ is the Euler's gamma function. Then, $A(\gamma)$
increases from 3.05 when $\gamma=2.0$ to 7.96 when $\gamma=1.3$.

An integration limit $\pi_{max}$ has to be chosen in Eq.~\ref{wdef} to
maximize the correlation signal. Indeed, one should avoid $\pi_{max}$
values that are too high, since they would mainly add noise to the
estimate of $w(r_p)$. On the other hand, scales that are too small,
comparable to the redshift uncertainties and to the pairwise
velocity dispersion (i.e. the dispersion in the distribution of the
relative velocities of source pairs), should also be avoided since
they would not allow the whole signal to be recovered.

The typical uncertainty in the redshift measurements
($\sigma_z<0.002)$ corresponds to comoving scales below $6.0\:h^{-1}$
Mpc at all redshifts. The pairwise velocity dispersion measured in the
local Universe ($500-600$ km $s^{-1}$; Marzke et al. \cite{marzke95},
Zehavi et al. \cite{zehavi02}) is expected to decrease by $\sim 15\%$
at $z\sim1.0$ (see e.g. the $\Lambda$CDM simulations by Kauffmann et
al. \cite{kauff99}), thus corresponding to $\sim 3\:h^{-1}$ Mpc. To
search for the best integration radius $\pi_{max}$, we measured
$w(r_p)$ for the XMM-COSMOS reference sample for different $\pi_{max}$
values ranging from 3 to $220\:h^{-1}$ Mpc. In Fig.~\ref{rov} (upper
panel), we show the increase of $w(r_p)$ with the integration radius
$\pi_{max}$ at those projected scales where most of the clustering
signal is coming from ($r_p=5-20\,h^{-1}$ Mpc). The $w(r_p)$ values
appear to converge for $\pi_{max}\gtrsim 40\,h^{-1}$ Mpc. Similarly,
the amplitude of the spatial correlation function $B=r_0^\gamma$
(Fig.~\ref{rov} bottom panel) or the amplitude of the projected
correlation function $C=A(\gamma)r_0^\gamma$ (not shown) is converging
for $\pi_{max}\gtrsim 40\,h^{-1}$ Mpc.\footnote{Since $A(\gamma)$
varies only by $\sim 16\%$ in the range of the measured slopes, $B$
and $C$ show almost exactly the same behaviour with $\pi_{max}$.} The
correlation length $r_{0}$ and slope $\gamma$ are strongly correlated:
when $r_0$ increases, $\gamma$ decreases. The correlation length
appears to reach a maximum at $\pi_{max}\sim 80\,h^{-1}$ Mpc, while
$\gamma$ is constant in the range $\pi_{max}=40-200\,h^{-1}$
Mpc. Based on these considerations, we adopt $\pi_{max}=40\:h^{-1}$
Mpc in the following analysis, which is the minimum $\pi_{max}$ value
at which the correlation function converges, and returns the
smaller errors on the best-fit correlation parameters $r_0,\gamma$ and
$r_0^\gamma$. We note that with this choice $r_0$ is smaller by
$\sim10\%$ than the maximum value measured at
$\pi_{max}\sim 80\,h^{-1}$ Mpc, but we do not try to correct for this
small bias.


To measure $\xi(r_p, \pi)$ we created random samples of sources in our
fields and measured the excess of pairs at separations $(r_p, \pi)$
with respect to the random distribution. We used the minimum variance
estimator proposed by Landy \& Szalay (\cite{ls93}), which is found to
have a nearly Poissonian variance and to outperform other popular
estimators, especially on large scales (e.g., see Kerscher et
al. 2000):

\begin{equation}
\xi(r_p, \pi)=\frac{[DD]-2[DR]+[RR]}{[RR]},
\end{equation}
\noindent
where [DD], [DR] and [RR] are the normalized data-data, data-random and
random-random pairs, i.e.

\begin{equation}
[DD]\equiv DD(r_p, \pi)\frac{n_r(n_r-1)}{n_d(n_d-1)}
\end{equation}

\begin{equation}
[DR]\equiv DR(r_p, \pi)\frac{(n_r-1)}{2 n_d}
\end{equation}

\begin{equation}
[RR]\equiv RR(r_p, \pi),
\end{equation}
\noindent
where $DD$, $DR$, and $RR$ are the number of data-data, data-random, and
random-random pairs at separations $r_p \pm \Delta r_p$ and $\pi \pm
\Delta \pi$, and $n_d$ and $n_r$ are the total number of sources in
the data and random sample, respectively.

\begin{figure*}[t]
\includegraphics[width=9cm]{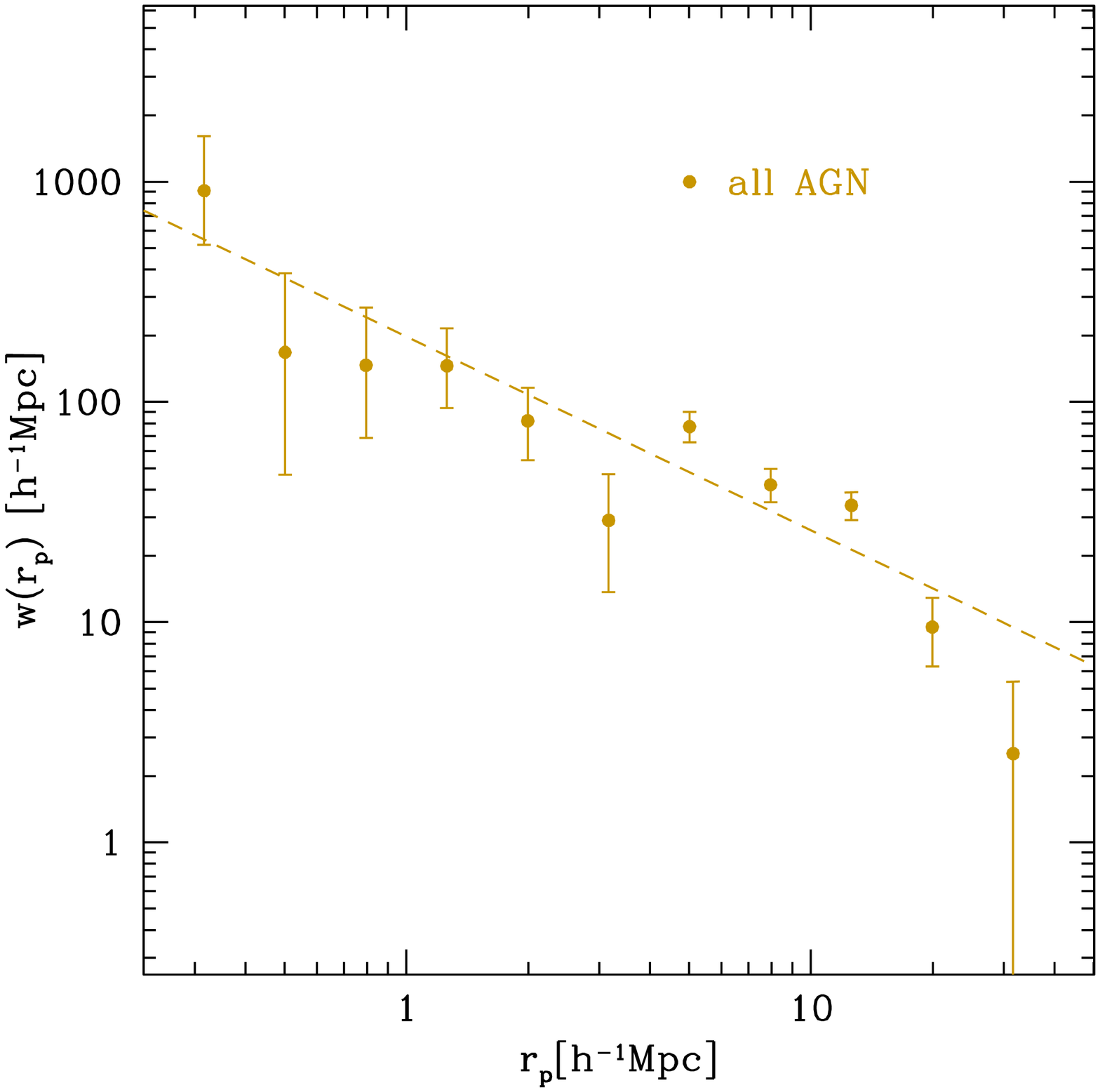}
\includegraphics[width=9cm]{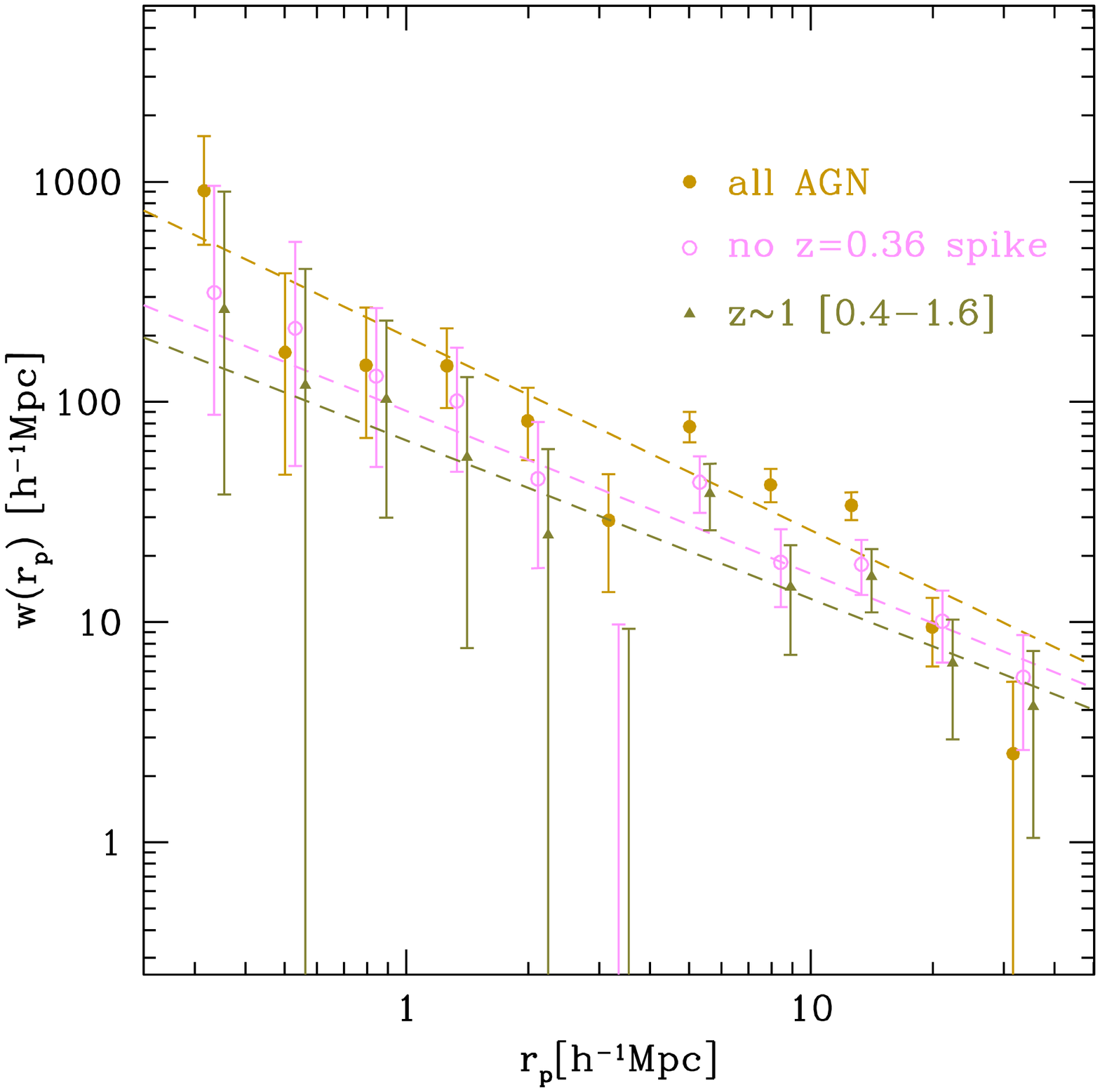}
\caption{{\it Left panel:} projected correlation function for the
XMM-COSMOS AGN reference sample (538 objects with $z=0.2-3.0$ and
$I_{AB}<23$, see Table~1). In this and in the following $panel$, errors
are $1\sigma$ Poisson confidence intervals and the best-fit power law
is shown as a dashed line. {\it Right panel:} projected correlation
function of the full sample computed including (open circles) and
excluding (filled circle) the 42 objects in the large-scale structure
at $z\sim 0.36$: it is evident that the $z\sim 0.36$ structure is
largely responsible for the signal excess on scales
$r_p\sim5-15\:h^{-1}$ Mpc. The projected correlation function for the
$z=0.4-1.6$ sample is also shown (filled triangles). For display
purposes, the $w(r_p)$ datapoints of the different samples have been
slightly shifted on the $r_p$ axis.}
\label{sall}
\end{figure*}

Since both the redshift and the coordinate $(\alpha, \delta)$
distributions of the identified sources are potentially affected by
observational biases, special care has to be taken in creating the
sample of random sources. This has been extensively discussed by Gilli
et al. (\cite{gilli05}) for the Chandra Msec Fields (see. e.g. their
Sect. 4.2) where similar problems have been encountered. In that paper
we showed that extracting the coordinates of the random sources from
the coordinate ensemble of the real sample and using the observed
redshifts to create a smoothed redshift distribution for the random
sample is a sufficiently accurate procedure. As in Gilli et
al. (\cite{gilli05}), we assumed a Gaussian smoothing length $\sigma_z
= 0.3$, which is a good compromise between scales that are too small,
which would suffer from local density variations, and those that are
too large, which would oversmooth the distribution. We nonetheless
verified that our results do not change significantly when using a
smoothing length in the range $\sigma_z = 0.2-0.4$. The smoothed
redshift distribution adopted for our simulations is shown in
Fig.~\ref{zdist}. The adopted procedure, if anything, would slightly
reduce the correlation signal, since it removes the effects of angular
clustering. Each random sample is built to contain more than 20000
objects.

We binned the source pairs in intervals of $\Delta{\rm log}\,r_p$=0.2
and measured $w(r_p)$ in each bin. The resulting datapoints were then
fitted by a power law of the form given in Eq.~\ref{wpow}, and the
best-fit parameters $\gamma$ and $r_0$ were determined via $\chi^2$
minimization. Given the small number of pairs that fall into some
bins (especially on the smallest scales), we used the formulae of
Gehrels (\cite{gehre86}) to estimate the 68\% confidence interval
(i.e. $1\sigma$ errorbars in Gaussian statistics).

\section{Safety checks and error estimates}

A possible concern related to the analysis methods presented in the
previous section is the random-sample generation. Indeed, placing the
random sources at the coordinates of the real sources completely
removes the contribution to the signal due to angular clustering. This
procedure could therefore underestimate the true correlation length.

We try to quantify this effect by considering a random sample
simulated according to the XMM-COSMOS sensitivity maps (see e.g Miyaji
et al. 2007 and Cappelluti et al. 2007). Briefly, each simulated
source is extracted from a reference input logN-logS, placed at random
in the XMM-COSMOS field, and kept in the random sample if its flux is
above the sensitivity map value at that position. It is evident that
this method is producing a random sample that only accounts for the
varying X-ray sensitivity along the COSMOS field, but does not account
for the positional biases related to the optical follow-up
program. 
The result of this test is that the measured correlation length
increases by $\sim 15\%$ with respect to the former case.  

An additional test was performed prompted by the X-ray flux
distribution of objects with spectroscopic redshift being different
from that of the total XMM sample. In particular, the fraction of
objects with spectroscopic redshift $Frac$ is constant (about 70\%)
for X-ray fluxes $f_{0.5-2 keV}>1.6\times 10^{-14}$ erg cm$^{-2}$
s$^{-1}$, while it decreases towards fainter fluxes, reaching 0.0 at
$f_{0.5-2 keV}<7\times10^{-16}$. Objects with spectroscopic redshifts
may therefore undersample the regions of maximum X-ray sensitivity, in
which the X-ray source density is higher, producing a more regular
distribution on the sky than the total XMM sample. We therefore
created a new random sample by first placing sources on the field
according to the X-ray sensitivity map as discussed in the previous
paragraph and then keeping only a fraction of them, with a
flux-dependent ``keeping" probability given by the observed relation
$Frac$ vs flux described above. When computing the projected
correlation function using this new random sample, we find a result
similar to what was obtained in the previous test, i.e. a $\sim 16\%$
higher $r_0$ value than obtained when placing random objects exactly
at the coordinates of real objects. 

Again, this new random sample unfortunately also does not fully
account for the positional biases related to the optical follow-up
programs. Indeed, given the very complex optical follow-up that
combines results from different programs, it is impossible for us to
estimate the correct selection function of our sample, but it is
likely that the selection of the masks used for optical spectroscopy,
which cover the COSMOS field unevenly, leaving some patches of the
field poorly covered, while covering other patches rather extensively,
is causing the main positional bias. This can be for instance
appreciated in Fig.~1, in which 3 of the 4 inner circles representing
Magellan IMACS pointings (16 pointings in total) have a higher density
of objects with spectroscopic redshift (cyan dots) than all the
remaining pointings. We therefore believe that the systematic upward
shift of $15-16\%$ in $r_0$ that we obtained with these tests is
likely to be an upper limit. Also, we note that, when performing error
analysis considering bootstrap errors (see next paragraph), a
difference of $15-16\%$ is within the total error budget. Given this
limited difference, we are confident that our results are not strongly
affected by the method used to generate the random source sample.

While many of the correlation function estimators used in the
literature have a variance substantially larger than Poisson (because
source pairs in general are not independent, i.e. the same objects
appear in more than one pair), the estimator used here was shown to
have a nearly Poissonian variance (Landy \& Szalay \cite{ls93}). It
has, however, to be noted that the Landy \& Szalay (\cite{ls93})
estimator was originally tested in the approximation of weak
clustering, so that Poisson errorbars may in our case underestimate
the true uncertainties. Bootstrap resampling has often been used to
estimate the uncertainties in the correlation function best-fit
parameters (e.g. Mo, Jing \& B\"orner \cite{mo92}), but this technique
may return an overestimate of the real uncertainties (Fisher et
al. 1994). We tested bootstrap errors by randomly extracting 100
samples of 538 sources each from our total sample, allowing for
repetitions. The {\it rms} in the distribution of the best-fit
correlation lengths and slopes is a factor of $\sim 2.8$ and $\sim 2$
greater than the Poisson errorbars, respectively. In the following we
simply quote $r_0$ and $\gamma$, together with their $1\sigma$
Poisson errors, bearing in mind that the most likely uncertainty is
about a factor of 2 higher.

\section {Results}

We first measured the projected correlation function $w(r_p)$ of all
the 538 spectroscopically identified sources in our XMM-COSMOS
reference sample ($I_{AB}<23$ and $z=0.2-3$) regardless of their
optical classification. The correlation function was measured on
projected scales $r_p=0.3-40\:h^{-1}$ Mpc. Here and in the following
samples, a simple power law is fit to the data, using standard
$\chi^2$ minimization techniques to get the best-fit parameters. The
best-fit correlation length and slope are found to be $r_0 = 8.65\pm
0.45\; h^{-1}$ Mpc $\gamma=1.88\pm 0.07$, respectively. Based on the
error on $r_0$ from this two-parameter fit, we estimate the
clustering signal to be detected at $\gtrsim 18 \sigma$ level, which
is the most significant clustering measurement to date for X-ray
selected AGN. The measured correlation length appears to be in good
agreement with what is estimated by Miyaji et al. (2007) on the first
pass on XMM-COSMOS through angular clustering and Limber's
inversion. As shown in Fig.~\ref{sall}a, a signal excess above the
power-law fit is observed in the scale range $r_p\sim 5-15\:h^{-1}$
Mpc, which deserves further investigation. As shown in
Fig.~\ref{zdist}, a significant fraction of XMM-COSMOS sources are
located within a large-scale structure at $z\sim 0.36$: overall, 42
objects are located at $z=0.34-0.38$, with sky coordinates distributed
all over the 2 deg$^2$ field.\footnote{Another $\sim 25$ X-ray
selected objects belonging to this structure are found using
photometric redshifts (see Fig.~\ref{iz} and Salvato et al. 2008).} We
verified that, when sources in the $z=0.34-0.38$ redshift range are
excluded, the signal excess at $r_p\sim 5-15\:h^{-1}$ Mpc actually
disappears\footnote{We note that, because of this excess due to
the $z\sim 0.36$ structure, a simple power law is a very poor fit to the
$w(r_p)$ of our full reference sample and of the $z<1$ AGN subsample
(see Sect. 5.3 and Table~1). In contrast, a simple power law
provides statistically acceptable fits for all the other subsamples
analyzed in this work.} (see Fig.~\ref{sall}b). This has the effect
of reducing the correlation length to $r_0=6.1\pm 0.8$ (see
Table~1). We also verified the effects of restricting the redshift
interval around the peak of the redshift distribution and computing
$w(r_p)$ in that interval. For XMM-COSMOS AGN in the range $z=0.4-1.6$,
we found a correlation length of $r_0\sim 5.2\pm1.0\:h^{-1}$ Mpc (see
Table~1). Similar results are found when the redshift interval is
restricted to $z=0.5-1.5$.

We are repeating our angular auto-correlation function analysis
using the full three-year data (55 pointings instead of the first-year
23 pointings used in Miyaji et al. 2007). We plan to use the accurate
($\approx 2\%$) photometric redshifts presented by Salvato et
al. (2008) to calculate the angular correlation function of XMM AGNs
selected by redshift. Preliminary results show $r_0=9.7\pm 2.2 h^{-1}$
Mpc for the $z>0$ sample, which is fully consistent with the
measurement presented in Miyaji et al. (2007), while the correlation
length is reduced to $r_0=8.9\pm 2.5 h^{-1}$ Mpc (1$\sigma$ errors)
for the sample obtained by excluding objects in the redshift range
$z=0.34-0.36$. These values are consistent within the errors with
those obtained with the present analysis, despite being systematically
higher. The angular analysis results, however, still vary with the
angular scale range used for fitting, as well as with error estimation
methods. A more detailed discussion of the comparison of the results
presented in this paper with those obtained through the study of the
angular auto-correlation function and cross-correlation function with
galaxies will be presented in a future paper (Miyaji et al., in
prep).

\begin{figure}
\includegraphics[width=8.5cm]{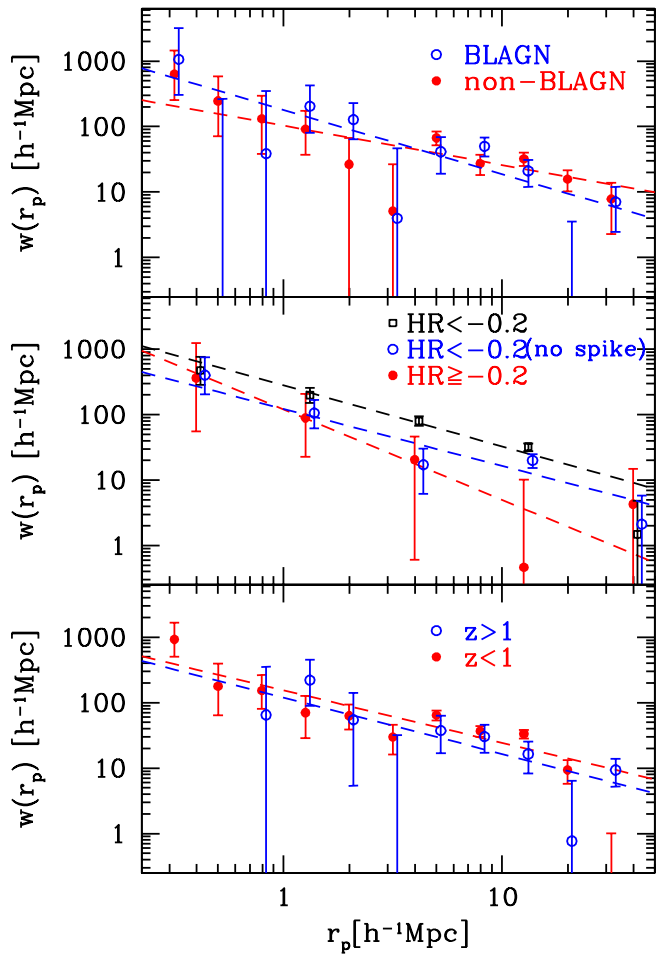}
\caption{Projected correlation function for different XMM-COSMOS AGN
subsamples.{\it Upper panel:} broad line AGN (open circles) vs
non-broad line AGN (filled circles). {\it Middle panel:} X-ray
unabsorbed ($HR<-0.2)$ AGN (open squares) vs X-ray absorbed
($HR\geq-0.2$) AGN (filled circles). Since the $z=0.36$ redshift
structure is mainly populated by X-ray unabsorbed objects (see
Fig.\ref{nhz} {\it right panel}), we also plotted the projected
correlation function obtained for X-ray unabsorbed AGN after removing
objects at $z\sim 0.36$ (open circles). {\it Lower panel:} AGN at
$z>1$ (open circles) vs AGN at $z<1$ (filled circles). In all panels
errors are $1\sigma$ Poisson confidence intervals and the best fit
power laws are shown as dashed lines.}
\label{xis}
\end{figure}

\subsection{AGN clustering as a function of optical type}

It is interesting to investigate the projected correlation function
for different source subsamples. For each subsample we placed the
sources of the random sample $only$ at the positions of the sources in
that subsample. The projected correlation function was then fitted
both leaving the slope free and fixing it to $\gamma=1.8$, the
standard value measured in most galaxy samples, which is also similar
to the slope of 1.88 measured for the total XMM-COSMOS sample. Fixing
the slope allows a more direct comparison between the correlation
lengths of the different subsamples when a two-parameter fit is poorly
constrained.


We investigated the clustering properties of sources optically
classified as broad line AGN (BLAGN) or non-BLAGN. The class of
non-BLAGN is admittedly a mixed bag, which may include obscured AGN,
weak unobscured AGN, whose optical emission is diluted by the host
galaxy light, and normal galaxies. The cut at $z>0.2$ (roughly
corresponding to $L_{0.5-2}>10^{41.5}$ erg s$^{-1}$), however, should
guarantee that the non-BLAGN sample is mostly populated by AGN in
which the absence of broad optical lines is solely due to nuclear
obscuration. Therefore, investigating the clustering properties of
BLAGN and non-BLAGN should be a proxy to investigate the clustering
properties of unobscured vs obscured AGN. For the sample of 305 BLAGN,
we measured a correlation correlation length of
$r_0\sim7.7\pm0.9\:h^{-1}$ Mpc and a slope of $\gamma\sim2.0\pm0.2$,
while for the 229 non-BLAGN we measured a similar correlation length,
$r_0\sim7.0\pm1.0\:h^{-1}$ Mpc, and a somewhat flatter slope
$\gamma\sim1.6\pm0.1$. The projected correlation functions of BLAGN
and non-BLAGN AGN are shown in Fig.~\ref{xis} {\it (upper panel)}.
It should be noted that a proper comparison between the clustering
properties of BLAGN and non-BLAGN should take possible redshift
effects into account, since BLAGN are generally observed at higher
redshift than non-BLAGN (median $z=1.5$ vs $z=0.7$, see Table~1). In
principle, the correlation length of a given AGN and galaxy population
is expected to change with redshift, being intimately related to the
evolution of the hosting dark matter halos, and one should therefore
compare source populations at the same redshift to establish whether
they reside in the same environment or not. We will return to this in
the discussion.

\begin{table*}
\begin{center}
\caption{Summary of best-fit clustering parameters.}
\begin{tabular}{lrcccccc}
\hline \hline
Sample& $N^a$& $z$ range& $\bar z^b$& ${\rm log}L_{0.5-10}^c$& $r_0$~~~~~~& $\gamma$& $r_0(\gamma=1.8)$\\
&&&&&[$h^{-1}$ Mpc]&&[$h^{-1}$ Mpc]\\
\hline
All AGN           & 538 & 0.2-3.0 & 0.98 & 43.8& $8.65^{+0.41}_{-0.48} $& $1.88^{+0.06}_{-0.07}$ & $8.39^{+0.41}_{-0.39}$ \\
No z=0.36 spike   & 496 & 0.2-3.0 & 1.03 & 43.9& $6.12^{+0.64}_{-0.89} $& $1.74^{+0.13}_{-0.14}$ & $6.32^{+0.53}_{-0.49}$ \\
\vspace{0.2truecm}
$z\sim 1$ AGN     & 349 & 0.4-1.6 & 0.94 & 43.7& $5.17^{+0.80}_{-1.14} $& $1.72^{+0.17}_{-0.18}$ & $5.44^{+0.62}_{-0.58}$ \\ 
BLAGN             & 305 & 0.2-3.0 & 1.45 & 44.3& $7.66^{+0.81}_{-1.04} $& $1.98^{+0.17}_{-0.18}$ & $7.03^{+0.96}_{-0.89}$ \\
\vspace{0.2truecm}
non-BLAGN         & 229 & 0.2-1.3 & 0.70 & 43.2& $7.03^{+0.87}_{-1.18} $& $1.60^{+0.13}_{-0.14}$ & $7.75^{+0.62}_{-0.59}$ \\
$HR<-0.2$         & 428 & 0.2-3.0 & 1.15 & 44.0& $9.56^{+0.42}_{-0.45} $& $1.98^{+0.08}_{-0.07}$ & $9.05^{+0.48}_{-0.45}$ \\
$HR<-0.2,\,z<1.3$ & 250 & 0.2-1.3 & 0.79 & 43.5& $9.96^{+0.48}_{-0.50}$& $1.93^{+0.07}_{-0.07}$ & $9.68^{+0.48}_{-0.50}$ \\ 
$HR<-0.2$,no spike& 218 & 0.2-1.3 & 0.79 & 43.5& $6.72^{+0.72}_{-0.88}$& $1.87^{+0.18}_{-0.18}$ & $6.56^{+0.75}_{-0.68}$ \\ 
\vspace{0.2truecm}
$HR\geq-0.2$      & 102 & 0.2-1.3 & 0.73 & 43.4& $5.07^{+1.58}_{-1.73} $& $2.38^{+0.85}_{-0.51}$ & $4.97^{+1.93}_{-1.49}$ \\
$z<1$             & 276 & 0.2-1.0 & 0.68 & 43.3& $7.97^{+0.48}_{-0.52} $& $1.80^{+0.08}_{-0.08}$ & $7.97^{+0.43}_{-0.41}$ \\
$z<1$ no spike    & 234 & 0.2-1.0 & 0.73 & 43.4& $5.16^{+0.71}_{-0.99} $& $1.81^{+0.20}_{-0.21}$ & $5.14^{+0.66}_{-0.61}$ \\
$z>1$             & 262 & 1.0-3.0 & 1.53 & 44.3& $6.68^{+1.19}_{-1.98} $& $1.86^{+0.27}_{-0.30}$ & $6.40^{+0.94}_{-0.86}$ \\
\hline		 	      			   
\end{tabular}

$^a$Number of objects in each sample. $^b$Median redshift. $^c$Median
X-ray luminosity in the 0.5-10 keV band in units of erg
$s^{-1}$. Errors are $1\sigma$ Poisson confidence levels. Bootstrap
errors are a factor of $\sim 2$ larger.

\end{center}
\end{table*}

\begin{figure*}[t]
\includegraphics[width=16cm]{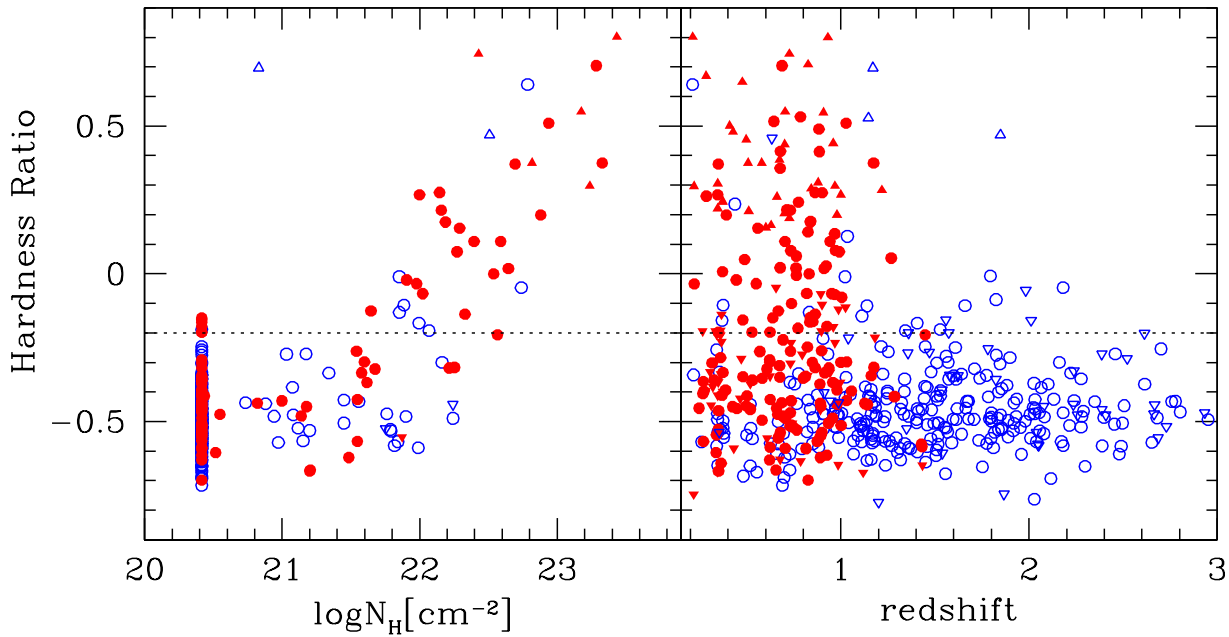}
\caption{{\it Left panel}. Column density $N_H$ vs hardness ratio
distribution for the 290 objects in our reference sample with more
than 200 photons in the 0.5-10 keV band, for which an accurate $N_H$
measurement can be performed through X-ray spectral fitting. Objects
that do not show absorption have been plotted at log$N_H$=20.4,
i.e. at the Galactic value. Broad line AGN and non-broad line AGN are
shown as blue open and red filled symbols, respectively. Circles refer
to objects detected in both soft and hard X-ray bands, for which an HR
value is directly measured. Upward (downward) pointing triangles
represent upper (lower) limits to the hardness ratios. It is evident
that most objects above the dotted line at $HR=-0.2$ are absorbed by
column densities $logN_H>21.6$. {\it Right panel}. Hardness ratio vs
redshift plot for the 538 sources in our reference sample. Symbols are
as in the {\it left panel}. The dotted line at $HR=-0.2$ marks the
adopted dividing line between X-ray absorbed and unabsorbed AGN.}

\label{nhz}
\end{figure*}

\subsection{AGN clustering as a function of X-ray absorption}

To investigate the clustering properties of obscured vs
unobscured AGN further, we also considered the column density measurements
obtained from the spectral analysis of XMM data. We considered here
the measurements performed by Mainieri et al. (2007 and in prep.), who
performed X-ray spectral fits for those objects with more than 100
counts in the 0.3-10 keV band and found absorption in excess of the
Galactic value in about 25\% of their sample (see details in Mainieri
et al. 2007). We considered here only objects with more than 200
counts in the 0.3-10 keV band, for which the determination of the
column density is more reliable. By matching the Mainieri et
al. objects with more than 200 counts with our reference sample, we end
up with 290 objects for which a column density has been estimated: 70
of these do show absorption in excess of the galactic value. We note
that 21 out of the 190 broad line AGN with measured column density
have absorption in excess of the Galactic value, which is consistent
with the 10\% fraction of X-ray absorbed broad line AGN found in other
X-ray selected samples (see e.g. Tozzi et al. 2006, Brusa et
al. 2003). A few words of caution should, however, be spent on these
sources. First, simulations run on input X-ray unabsorbed spectra show
that, especially for sources at high redshift and relatively low
photon statistics, the spectral fit may return spurious positive
values for the absorption (see e.g. Tozzi et al. 2006). Second, the
fraction of X-ray absorbed sources might be related to the
significance threshold used to assess the presence of X-ray
absorption. In particular, Mainieri et al. consider an X-ray source as
absorbed if the addition of a photoelectric cut off in the spectral fit
improves it a a level of more than 90\% as assessed by an F-test. One
would then expect that, in about 7 out of 70 absorbed sources (either
BLAGN or non-BLAGN), the measured column density is spurious. At any
rate, results do not change significantly if we include those 21
candidate X-ray absorbed BLAGN in the total X-ray absorbed sample or
not.

We first verified that the projected correlation function of the 290
objects with X-ray spectroscopy is consistent with that of our full
reference sample and then tried to measure the projected correlation
function for absorbed and unabsorbed sources
separately. Unfortunately, the small number statistics prevent us from
getting a significant clustering signal for the 70 X-ray absorbed AGN,
while for the 220 X-ray unabsorbed objects we found a correlation
length of $r_0=10.6\pm0.9$, somewhat higher than that measured for the
subsample of BLAGN. When restricting the analysis to the redshift
range z=0.4-1.6, i.e. around the peak of the redshift distribution,
which also excludes the redshift structure at z=0.36, the correlation
length of the X-ray unabsorbed objects decreases to $r_0=6.2\pm1.4$,
consistent with what is measured for the full reference sample in the
same redshift interval.

To overcome the limitations due to the small size of the sample of
objects with reliable $N_H$ measurements, we tried to calibrate a
relation between the X-ray column density $N_H$ and the hardness ratio
(HR), defined as the difference between the source X-ray photons
detected in the 2-10 keV band and those detected in the 0.5-2 keV
band, normalized to the sum of the photons in the two
bands.\footnote{Upper (lower) limits to the hardness ratio for objects
that are not detected in the soft (hard) band have been computed
assigning 40 counts in the non-detection band to each source. The
distribution of the counts of the detected sources indeed shows a
turnover at this value, which therefore appears as the average count
threshold for source detection.} The distribution of the column
density vs hardness ratio for objects with more than 200 counts is
shown in Fig.~\ref{nhz} ({\it left panel}): most of the objects with
$HR>-0.2$ do show absorption in excess of log$N_H=21.6$, therefore we
adopt a rough threshold at $HR=-0.2$ to divide X-ray absorbed from
X-ray unabsorbed AGN (see also Hasinger 2008 and Fig.~11 in Mainieri
et al. 2007). The HR distribution as a function of redshift for the
538 objects in the reference sample is shown in Fig.~\ref{nhz} ({\it
right panel}). Most BLAGN fall below the HR=--0.2 line, while
non-BLAGN AGN do show higher HR values on average. The poorly
populated upper-right corner of the figure, i.e. the high-HR - high-z
region, suffers from obvious selection effects due to i) the bias
against faint (distant and absorbed) magnitude targets in the optical
spectroscopy follow-up (see also Fig.~4 in Brusa et al. 2007) and ii)
the K-correction effects that make high-redshift absorbed spectra to
appear softer in the X-ray bandpass (i.e. lower HR values). It is
noted that BLAGN form a sort of horizontal sequence at $HR\sim-0.5$,
which is indeed the hardness ratio value expected by a canonical power
law spectrum with photon index $\Gamma=1.7$ and no absorption. Since
we consider all objects with $HR<-0.2$ as unabsorbed sources, the
adopted cut conservatively accounts for any dispersion in the photon
index distribution of BLAGN. We measured $w(r_p)$ for absorbed and
unabsorbed objects separately. At projected scales below
$r_p\sim1\;h^{-1}$ Mpc, absorbed and unabsorbed AGN are similarly
correlated, while absorbed AGN appear less correlated on larger
scales. This results in absorbed AGN formally having a lower
correlation length ($r_0= 5.1\pm1.7\;vs\;9.6\pm0.4\;h^{-1}$ Mpc) and a
steeper slope ($\gamma= 2.4\pm0.7\;vs\;2.0\pm0.1$) than unabsorbed
objects. The projected correlation function for unabsorbed AGN does
not change significantly if we restrict the analysis to the redshift
range 0.2-1.3, i.e. the same range as used for absorbed AGN (see
Table~1). The projected correlation function of unabsorbed and
absorbed AGN in the same z=0.2-1.3 redshift interval are shown in
Fig.~\ref{xis} ({\it middle panel}). The larger correlation length
measured for unabsorbed objects is essentially due to most objects in
the z=0.36 structure having $HR<-0.2$ (see Fig.~\ref{nhz}, {\it right
panel}). Indeed, when removing this structure, the correlation length
for X-ray unabsorbed AGN decreases to $r_0=6.7\pm0.8$ (see
Fig.~\ref{xis}, {\it middle panel}), which, given the large errorbars,
is not significantly different from that of X-ray absorbed objects. To
summarize, from our analysis we cannot claim that X-ray absorbed and
X-ray unabsorbed AGN possess different clustering properties.

\subsection{AGN clustering as a function of redshift}

The study of AGN clustering as a function of redshift provides several
pieces of information about the formation and evolution of the AGN
population.

Because of the limited sample size, we simply split the XMM-COSMOS AGN
sample in two subsamples of objects below and above redshift 1. The
correlation functions of the 276 AGN at $z<1$ and of the 262 AGN at
$z>1$ are shown in Fig.~\ref{xis} {\it lower panel}. The best-fit
correlation parameters for objects at $z<1$ are found to be $r_0\sim
8.0\pm0.5\:h^{-1}$ Mpc and $\gamma\sim1.8\pm0.1$, while for objects at
$z>1$ the best-fit parameters are $r_0\sim 6.7\pm1.6\:h^{-1}$ Mpc and
$\gamma\sim1.9\pm0.3$. When removing the $z=0.36$ structure, the
correlation length of objects at $z<1$ decreases to $r_0\sim
5.2\pm0.8\:h^{-1}$ Mpc (see Table~1). In Sect. 6.3. we discuss the
correlation lengths of the various XMM-COSMOS redshift subsamples as
compared to those of other optical and X-ray selected samples at
different redshifts.

\subsection{AGN clustering as a function of luminosity}

We finally investigated the dependence of the AGN clustering
parameters on the X-ray luminosity, since this may reveal whether
objects shining with different luminosities reside in dark matter
halos with different masses, hence constraining the distribution of
the AGN Eddington ratios (see eg. Lidz et al. 2006, Marulli et
al. 2008 and the discussion in Sect. 6.3). Again, because of the
limited size of the sample, we simply divided it into two almost
equally populated subsamples, the dividing line being at $L_{0.5-10
keV}=10^{44}$ erg s$^{-1}$ (corresponding to $L_{0.5-2
keV}\sim10^{43.5}$ erg s$^{-1}$ for a typical AGN X-ray spectrum). As
shown in Fig.~\ref{lxz}, splitting the sample at this luminosity is
almost equal to splitting the sample at a redshift of $z\sim
1$. Indeed, when computing the clustering parameters of the higher
(lower) luminosity sample, these are very similar to the $z>1$ ($z<1$)
sample, and therefore are not reported here.

\section{Discussion} \label{discussion}

\subsection{Comparison with other $z\sim 1$ AGN and galaxy samples}

The galaxy and AGN census in the $z\sim 1$ Universe has been recently
enlarged by a number of surveys with different areas and
sensitivities, which allowed investigation of the spatial distribution
of different populations. The comparison between the clustering
properties of AGN and galaxies allows to first approximation to infer
which galaxy population is hosting any given AGN population, under the
simple hypothesis that AGN activity at a given redshift is randomly
sampling the host galaxy population. The comparison between $z\sim1$
AGN samples obtained from surveys with different sensitivities may
also reveal any dependence of AGN clustering on luminosity. As far as
X-ray selected AGN are concerned, a correlation length of
$r_0=5.7^{+0.8}_{-1.5}\,h^{-1}$ Mpc has been measured for $\sim 230$
objects in the 0.4 deg$^2$ CLASXS survey (Yang et al. 2006). Objects
in the CLASXS have redshift and luminosity distributions very similar
to those of our sample, and therefore they should trace the AGN
population sampled by XMM-COSMOS almost exactly. To check whether the
different techniques used for the clustering analysis may introduce
significant differences, we analyzed the CLASXS sample using the same
techniques as were used in this work finding best-fit clustering
parameters in very good agreement with the Yang et al. (2006)
values. The difference between the $r_0$ values measured in XMM-COSMOS
and in CLASXS full samples (8.6 vs 5.7 $h^{-1}$ Mpc, respectively)
therefore appears to be inherent to the two fields considered. While a
prominent redshift spike is observed at z=0.36 in XMM-COSMOS, no such
similar structures are found in the CLASXS field. Indeed, when
removing the structure at $z=0.36$, the correlation length of
XMM-COSMOS AGN decreases to $\sim6.3\,h^{-1}$ Mpc, in good agreement
with the value measured in CLASXS. Moreover, when restricting the
analysis to XMM-COSMOS AGN in the redshift range $z=0.4-1.6$, the
correlation length ($\sim5.2\,h^{-1}$ Mpc) is very similar to what is
measured in CLASXS. One therefore may wonder about the frequency with
which prominent large-scale structures are sampled in X-ray surveys of
different sky areas, i.e. the effects of cosmic variance. Indeed,
based on simulated galaxy mock catalogs over 2 deg$^2$ fields
(Kitzbichler \& White 2006), some evidence exists that the COSMOS
field has some excess of structures with respect to the average.

The correlation length of XMM-COSMOS AGN can be compared to that of
different galaxy populations at $z\sim 1$. Coil et al.\ (2004) find
$r_0=3.2\pm0.5\,h^{-1}$ Mpc for emission line galaxies in the DEEP2
survey, while Meneux et al.\ (2006) find $r_0=2.5\pm0.4\,h^{-1}$ Mpc
for star-forming blue galaxies in the Vimos-VLT Deep Survey (VVDS, Le
Fevre et al. 2004). The populations of red absorption-line galaxies
in the same surveys have instead larger correlation lengths:
$r_0\sim6.6\,h^{-1}$ Mpc for absorption line galaxies in the DEEP2
(Coil et al. 2004) and $r_0=4.8\pm0.9\,h^{-1}$ Mpc for red, early type
galaxies in the VVDS (Meneux et al. 2006). Recently, a correlation
length as large as $r_0=5.1\pm0.8\,h^{-1}$ Mpc has been measured for
luminous infrared galaxies (LIRG, $L_{IR}>10^{11}\,L_{\odot}$) at
$z\sim 0.8$, which are forming stars at high rates ($SFR>17 M_{\odot}$
yr$^{-1}$; see Gilli et al. 2007b). Since at $z\sim 1$ star formation
is closely related to galaxy mass (Noeske et al. 2007, Elbaz et
al. 2007), even LIRGs, as well as $z\sim 1$ early type galaxies, are
massive objects with stellar mass $M_\star\gtrsim 3\times 10^{10} \;
M_{\odot}$. The fact that XMM-COSMOS AGN show similar correlation
length to these systems (see Fig.~\ref{nr0}), suggests that, similar
to what is observed at $z=0$ (Kauffmann et al. 2004) and at $z\sim 2$
(Daddi et al. 2007), at $z\sim 1$ nuclear activity is hosted by
the more massive galaxies (see also Georgakakis et al. 2007). This is
in good agreement with the analysis by Silverman et al. (2008b), who
investigated the occurrence of nuclear activity on a sample of $\sim
8000$ galaxies selected from the zCOSMOS spectroscopic catalog,
finding that the fraction of galaxies hosting an AGN increases towards
large stellar masses at $z\lesssim 1.0$. In particular, most ($\sim
80\%$) AGN at $z\sim 1$ reside in galaxies with stellar mass $M_\star
> 3\times 10^{10} \; M_{\odot}$, in agreement with the conclusions
from our clustering analysis.




\subsection{The connection with dark matter halos}
\label{halomass}

While on small scales, comparable to the dimensions of dark matter
halos, AGN and galaxy clustering are difficult to predict because of
merging and interactions that can trigger a number of physical
processes, on larger scales (e.g., $>1\;h^{-1}$ Mpc), where
interactions are rare, the AGN correlation function should follow that
of the hosting dark matter halos.

An interesting consequence is that one can estimate the masses of the
typical halos hosting an AGN population by simply comparing their
clustering level. According to the standard $\Lambda$CDM hierarchical
scenario, dark matter halos of different mass cluster differently,
with the more massive halos more clustered for any given epoch,
and it is straightforward to compute the correlation function for
halos above a given mass threshold. It is worth noting that, since less
massive halos are more abundant, the correlation function of halos
above a given mass threshold is very similar to the clustering of
halos with mass close to that threshold. Also, it is important to note
that, as far as our measurements are concerned, the best-fit clustering
parameters are obtained from datapoints mostly on large scales
($r_p>1\;h^{-1}$ Mpc; see Fig.~\ref{sall}). Therefore the measured
$r_0$ and $\gamma$ values are essentially due to the clustering signal
on large scales, where the AGN correlation function follows that of
the dark matter, allowing a meaningful comparison with the clustering
expected for dark matter halos.

\begin{figure}
\includegraphics[width=9cm]{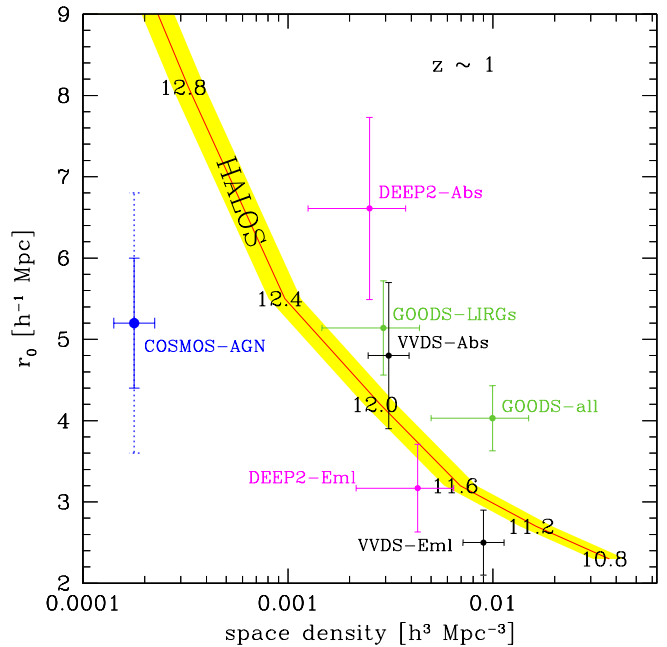}
\caption{Space density vs correlation length for the XMM-COSMOS AGN
compared to that of other AGN and galaxy populations at $z\sim 1$, as
labeled. The trend for the Millennium dark matter halos at $z\sim1$
above different mass thresholds is also shown as a shaded region. More
massive halos (log of the threshold mass is labeled) are less abundant
and more clustered than less massive ones. The XMM-COSMOS AGN
datapoint (big filled circle) refers to the z=0.4-1.6 sample. Solid
and dotted errorbars correspond to Poissonian and bootstrap
uncertainties, respectively.}
\label{nr0}
\end{figure}

We considered the dark matter halo catalogs available for the
Millennium simulation\footnote{see {\tt
http://www.mpa-garching.mpg.de/millennium}.}(Springel et
al. 2005). Halo catalogs are available at different time steps along
the simulation. Here we considered those at $z\sim1$ (parameter {\tt
stepnum=41} in the simulation). In total there are about $1.6\times
10^7$ halos with mass above $10^{10}\;h^{-1}\;M_{\odot}$ in a cubic
volume of 500$\;h^{-1}$ Mpc on a side. We computed the correlation
function and the space density of halos above 7 mass thresholds
ranging from log($M/M_{\odot}$)=10.8 to 13.2 in steps of 0.4. Here we
use as halo mass estimator the simulation parameter {\tt m\_Crit200},
defined as the mass within the radius where the integrated halo
overdensity is 200 times the critical density of the simulation. The
halo correlation length was estimated by fitting with a power law
the halo correlation functions on scales above $r=1\;h^{-1}$ Mpc. The
results are shown in Fig.~\ref{nr0}, where it is clear that
more massive halos are more clustered and less numerous.

We computed the space density of AGN in XMM-COSMOS as expected from
published X-ray luminosity functions (see details in Sect. 6.5) and
compared the $r_0$ and density values of our population with those of
other AGN and galaxy populations at $z\sim1$ and with those of dark
matter halos at $z\sim1$ as computed above. We considered the
XMM-COSMOS AGN in the redshift range $z=0.4-1.6$, i.e. around the peak
of the selection function and excluding the redshift structure at
$z=0.36$. The comparison is shown in Fig.~\ref{nr0}.


By comparing the halo and the galaxy $r_0$ values, XMM-COSMOS AGN
appear to be hosted by halos with masses $\gtrsim 2.5\times
10^{12}\;M_{\odot}$, similar to absorption line galaxies and LIRGs,
which indeed show similar correlation lengths. However, while
absorption line galaxies and LIRGs appear to be more abundant than the
hosting halos (with an average of 2-4 such galaxies per halo),
XMM-COSMOS AGN appear to be a factor of $\gtrsim 5$ less abundant than
the hosting halos, suggesting that nuclear activity is present in
about $\sim 15-20\%$ of halos of that mass. Considerations about the
duty cycle and lifetimes of XMM-COSMOS AGN will be presented in
Sect. \ref{life}.

\begin{figure}
\includegraphics[width=9cm]{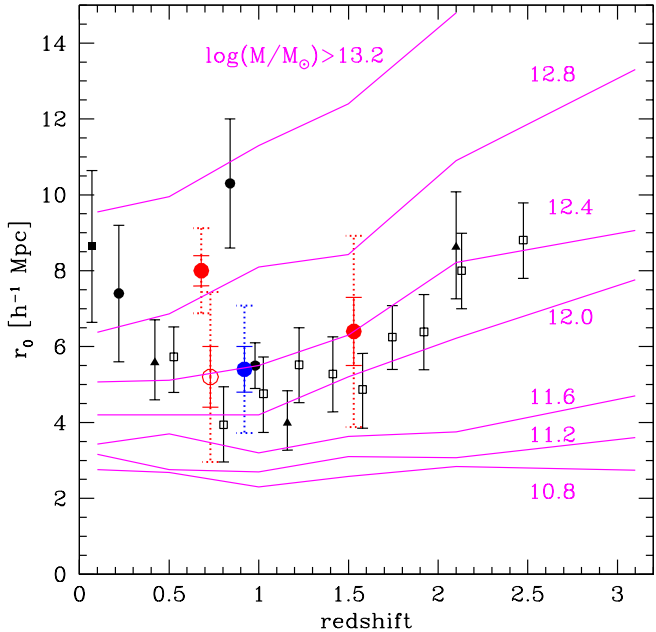}
\caption{Correlation length vs redshift for XMM-COSMOS AGN compared to
that of other AGN samples. The big blue filled circle refers to the
z=0.4-1.6 XMM-COSMOS sample. Big red filled circles refer to the $z<1$
and $z>1$ samples. The big open circle refers to the $z<1$ sample when
excluding the z=0.36 spike. Solid and dotted errorbars correspond to
Poissonian and bootstrap uncertainties, respectively. The filled
square at $z\sim0.1$ refers to the AERQS QSO sample (Grazian et
al. 2004). Filled circles at $z\sim0.2, 0.7, 0.9$ refer to the NEP
(Mullis et al. 2004), CDFS and CDFN samples (Gilli et al. 2005),
respectively. Filled triangles refer to the CLASXS sample (see eg Yang
et al. 2006). Open squares refer to optically selected QSOs in the 2QZ
sample (Croom et al. 2005). The solid curves show the correlation
length vs redshift relations expected for dark matter halos in the
Millennium simulation above different mass thresholds as labeled.}
\label{roz}
\end{figure}

\subsection{Evolution of AGN clustering}

To investigate the evolution of the AGN clustering properties with
redshift we combined the results from XMM-COSMOS with recent findings
from other X-ray and optical surveys. When necessary, the results were
corrected to the cosmology adopted here. The values of the correlation
lengths reported in this section were usually calculated by fixing
$\gamma$ to 1.8, therefore allowing a consistent comparison. When
different slopes were measured/adopted, we discuss the case and verify
the effects of assuming $\gamma=1.8$. Only results from spatial
clustering analysis are considered.

As for the X-ray surveys we considered the results from the ROSAT NEP
survey and from the Chandra Msec fields. In the NEP survey, Mullis et
al. (\cite{mullis04}) found a correlation length of
$r_0\sim7.4\pm1.8\:h^{-1}$ Mpc on scales of $5-60\:h^{-1}$ Mpc for
source pairs at a median redshift $\bar z = 0.22$. In the CDFS and
CDFN, the correlation length measured by Gilli et al. \cite{gilli05} is
$r_0=10.3\pm 1.7\:h^{-1}$ Mpc and $r_0=5.5\pm 0.6\:h^{-1}$ Mpc,
respectively. Although the best-fit slopes in the Chandra Msec fields
are rather flat ($\gamma=1.3-1.5$), the best-fit correlation lengths
increase by only $\sim15\%$ if the slope is fixed to 1.8 (Gilli et
al. \cite{gilli05}).

As for optically selected AGN, we considered the results of Croom et
al. (\cite{croom05}) based on more than 20000 objects in the final
catalog of the 2dF QSO Redshift Survey (2QZ), where the QSO
correlation length is found to increase significantly from $z\sim 1$
to $z\sim 2.5$, and very flat slopes ($\gamma\sim1.1-1.2$) have been
measured. As discussed by Croom et al. (\cite{croom05}), these flat
slopes stem from redshift-space distortions that are relevant when the
correlation function is measured down to small scales in redshift
rather than in real space. The real-space clustering for the total 2QZ
sample has instead been measured by Da Angela et al. (\cite{daa05})
via the projected correlation function. On the same scales as
considered by Croom et al. (\cite{croom05}) and when approximating
$\xi(r)$ by a single power law, they found that, while the slope of
the total 2QZ sample steepens significantly from $\gamma=1.20\pm0.10$
to $\gamma=1.85\pm0.13$, the correlation length only marginally
decreases by 10\% (from $r_0=5.5\pm0.5\:h^{-1}$ Mpc to
$r_0=5.0\pm0.5\:h^{-1}$ Mpc). In the following we therefore simply
consider the values as measured by Croom et al. (\cite{croom05}) for
the 2QZ correlation lengths in different redshift bins. These results
are consistent with those obtained by Porciani et al. (2004) using 2dF
QSOs in a narrower redshift range. In the local Universe ($z\sim 0.07$),
the clustering of bright optical QSOs ($B<15$ mag) has been recently
determined by Grazian et al. (\cite{grazi04}) by means of the
Asiago-ESO/RASS QSO survey (AERQS). These authors measured $r_0=8.6
\pm 2.0 \:h^{-1}$ Mpc at a median redshift of $z\sim0.1$ on comoving
scales $1-30 \: h^{-1}$ Mpc by fixing the correlation slope to
$\gamma=1.56$. Given the above considerations for the Chandra Msec
fields and the 2QZ and given the rather large uncertainties we have to
deal with, we consider the value quoted by Grazian et al. (2004) as if
obtained by fixing $\gamma$ to 1.8. All the measurements discussed above
are shown in Fig.~\ref{roz}.


Unfortunately, it is not possible to perform a completely unbiased
comparison between the various samples because different redshifts
generally sample different luminosities, and AGN clustering may be a
function of AGN luminosity if the latter correlates with the mass of
the hosting dark halo (e.g. Kauffmann \& Haehnelt \cite{kauff02}). 

In Fig.~\ref{roz}, sources at $z<0.3$ and $z>2$ appear to be the most
clustered ones, and these also correspond to the most luminous AGN. At
$z<0.3$ the median 0.5-10 keV luminosity of the AGN in the AERQS and
NEP samples is about log$L_x=44.4$ (see Mullis et al. \cite{mullis04}
and Gilli et al. \cite{gilli05}). At $z>2$ the median absolute B-band
luminosity of the 2QZ QSOs corresponds to a median 0.5-10 keV
luminosity of log$L_x=44.7$ (assuming a standard QSO SED, e.g. Elvis
et al. \cite{elvis94}). The less clustered sources are found at $z\sim
1$, but these have lower luminosities (log$L_x=43-44$). In general, a
clear dependence of clustering amplitude on AGN luminosity has not
been observed yet. On the contrary, the available evidence, if any,
points towards a similar clustering for sources at the same redshift
but with different luminosities. Croom et al. (\cite{croom05}) and
Porciani \& Norberg (2006) could not find any significant evidence of
luminosity dependent clustering in the 2QZ. From a cross-correlation
analysis between galaxies and AGN, Adelberger \& Steidel (\cite{as05})
claim that AGN at $z\sim 2$ cluster similarly within a 10 mag
luminosity range. From a theoretical point of view, one would expect
little clustering dependence on the observed AGN luminosity if this is
not directly related to the host halo mass; i.e., if, at any given
redshift, objects that reside in halos within a narrow mass range have
very different luminosities (eg. Lidz et al. 2006). For instance, even
assuming a dependence of black hole (and host galaxy) mass on the
hosting halo mass, a wide spread in the distribution in the Eddington
ratios would make BH of similar masses radiate at very different
luminosities. Indeed, although the average Eddington ratio of SDSS
QSOs has been shown to increase towards high luminosities, the spread
in the distribution is wide (McLure \& Dunlop 2004). Moreover, it has
been recently suggested (Gavignaud et al. 2008) that the dispersion in
the black hole mass--luminosity increases even more for lower AGN
luminosities. Finally, in the local Universe, Constantin \& Vogeley
(2006) find that low-luminosity LINERs are more clustered than higher
luminosity Seyfert galaxies, showing that the relation between AGN
luminosity and clustering may even be reversed for low-luminosity AGN
with respect to the expectations based on a monotonically increasing
relation between luminosity and black hole mass. Large statistical
samples, beyond the reach of the data presented in this work, are
needed to firmly establish any dependence of AGN clustering on
luminosity (see eg. Porciani \& Norberg 2006).

Overall, when removing the redshift structure at z=0.36, the
clustering of XMM-COSMOS AGN appear in good agreement with what is
measured for optical and X-ray selected AGN at different redshifts. A
larger correlation length is instead found for objects at $z<1$ if the
redshift structure is not removed.

To interpret our clustering measurements at different redshifts, we
considered the halo catalogs in the Millennium simulation and computed
their correlation function above different halo mass thresholds and at
different redshifts. We essentially repeated the computation presented
in Sect. 6.2 for halos at $z=0.1, 0.5, 1.0, 1.5, 2.0,$ and 3.0. The $r_0$
vs redshift curves for halos above different mass thresholds are shown
in Fig.~\ref{roz}. Both our $z\sim 1$ and $z>1$ AGN samples seem to be
hosted by halos with mass above $log(M/M_{\odot})>12.4$. For
XMM-COSMOS AGN at $z<1$ the minimum mass of the host halos varies from
12.4 to 12.8 depending on whether the z=0.36 structure is excluded or
included from the computation of $r_0$. 

\begin{figure}
\includegraphics[width=9cm]{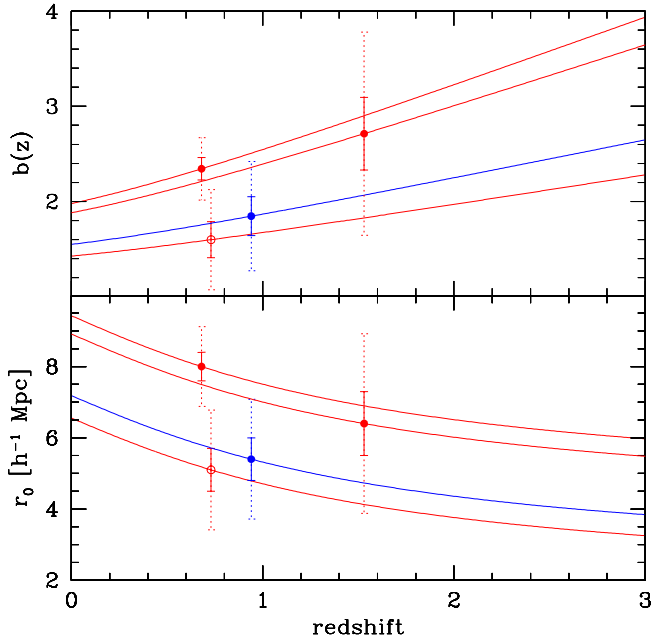}
\caption{Expected redshift evolution of the bias and correlation
length of different XMM-COSMOS AGN samples according to a {\it
conserving scenario}. The evolution curves are normalized to the
observed datapoints. Blue filled circle: $z=0.4-1.6$ sample. Red
circles: $z<1$ and $z>1$ samples. The open circle shows the result for
the $z<1$ sample when the $z=0.36$ structure is excluded. Solid and
dotted errorbars correspond to Poissonian and bootstrap uncertainties,
respectively.}
\label{broz}
\end{figure}

\subsection{Descendants of $z\sim1$ AGN}\label{evol}

As shown in the previous sections, under simple assumptions, it is
possible to use the spatial clustering of an extragalactic source
population measured at a given epoch to estimate the typical dark
matter halos in which these objects reside. Furthermore, it is also
possible to estimate their past and future history by following the
halo evolution in the cosmological density field. A useful quantity
for such analyses is the bias factor, defined as
$b^2(r,z,M)=\xi_A(r,z,M)/\xi_m(r,z)$, where $\xi_A(r,z,M)$ and
$\xi_m(r,z)$ are the correlation function of the considered AGN or
galaxy population and that of dark matter, respectively. In general
the bias parameter can be a function of scale $r$, redshift $z$, and
object mass $M$. For simplicity we adopt the following definition
here:
\begin{equation}
b^2(z)=\xi_A(8,z)/\xi_m(8,z)
\label{bb}
\end{equation}
in which $\xi_A(8,z)$ and $\xi_m(8,z)$, are the galaxy and dark matter
correlation functions evaluated at 8 $h^{-1}$ Mpc, respectively. The
AGN correlation function has been measured directly in this work,
while the dark matter correlation function can be estimated using the
following relation (e.g., Peebles 1980):
\begin{equation}
\xi_m(8,z)=\sigma_8^2(z)/J_2
\end{equation}
where $J_2=72/[(3-\gamma)(4-\gamma)(6-\gamma)2^\gamma]$, and
$\sigma_8^2(z)$ is the dark matter mass variance in spheres of 8
$h^{-1}$ Mpc comoving radius, which evolves as
$\sigma_8(z)=\sigma_8(0)D(z)$.
Also, $D(z)$ is the linear growth factor of perturbations, while
$\sigma_8(0)$ is the $rms$ dark matter fluctuation at the present
time, which we fix to $\sigma_8=0.8$ in agreement with the recent
results from WMAP3 (Spergel et al.\ 2007).\footnote{Only small
differences arise if we assume $\sigma_8=0.9$ as in the Millennium
simulation. The other relevant cosmological parameters assumed in this
work are the same as in the Millennium simulation.} While in an
Einstein - De Sitter cosmology the linear growth of perturbations is
simply described by $D_{EdS}(z)=(1+z)^{-1}$, the growth of
perturbations is slower in a $\Lambda$-dominated cosmology. We
consider here the so-called growth suppression factor
$g(z)=D(z)/D_{EdS}(z)$ as approximated analytically by Carroll, Press
\& Turner (1992). The resulting bias for the AGN in the XMM-COSMOS
field is $b(1.0)=2.0\pm0.2$.

Once the bias of the AGN population at its median redshift has been
estimated using the above relations, it is possible to follow the time
evolution of the bias with models that rely on simple assumptions.
Two popular scenarios that encompass two extreme hypotheses are the
{\it conserving model}, in which objects do not merge at all (Nusser
\& Davis 1994, Fry 1996) and the {\it merging model} in which objects
merge continuously (e.g. Moscardini et al. 1998). In the first
hypothesis the number of objects is conserved in time, and galaxies
behave as test particles whose spatial distribution simply evolves
with time under the gravitational pull of growing dark matter
structures. In the second hypothesis, object merging follows the
continuous merging of the hosting dark matter halos, in such a way
that only those objects and halos which have just merged are
observable at any given epoch.

From an observational point of view, the fraction of galaxies in
mergers appears to be a very debated issue. Recent works suggest that
close galaxy pairs (merger candidates) are a strong function of
redshift, evolving as $(1+z)^{3-4}$ (Kartaltepe et al. 2007, Kampzyck
et al. 2007). By extrapolating the current estimates, at $z\sim 2$
about 50\% of luminous galaxies are expected to be found in close
pairs/mergers. However, at $z\sim 1$ the fraction of galaxies in close
pairs is still $\sim 8\%$ and decreases to $\sim 0.1\%$ at $z\sim
0.1$.
%
In the following we will consider the non merging {\it conserving
model} as a fairly adequate representation of the bias evolution of
$z\sim 1$ XMM-COSMOS AGN towards lower redshifts; i.e., it will be used
to estimate the likely descendants of XMM-COSMOS AGN. On the contrary,
since merging is expected to be significant towards higher redshift, we
will not try to estimate their high-z progenitors.

In the {\it galaxy conserving model}, the bias evolution can be
approximated by
\begin{equation}
b(z)=1+[b(0)-1]/D(z)
\end{equation}
where $b(0)$ is the population bias at $z=0$ (Nusser \& Davis 1994,
Fry 1996, Moscardini et al. 1998). Once $b(z)$ is determined, the
evolution of $\xi_A(8,z)$ and hence of $r_0(z)$ can be obtained by
inverting Eq.~\ref{bb}. A value of $\gamma\sim 1.8$ for the slope is
assumed in the above relations. Little difference arises when using
$\gamma=1.9$.

In Fig.~\ref{broz} we show the evolution of $b(z)$ and $r_0(z)$ for
various XMM-COSMOS AGN samples, including the $z=0.4-1.6$ sample, the
sample at $z>1$ and the sample at $z<1$ with or without the structure
at $z=0.36$. By $z=0$ the correlation length of XMM-COSMOS AGN should
evolve to $r_0\gtrsim 6\,h^{-1}$ Mpc, which is typical of passive,
early type galaxies in the local Universe (Colless et al. 2001, Zehavi
et al. 2004). The correlation slope of the local early type population
$\gamma\sim1.8-2.0$ also appears consistent with that of XMM-COSMOS
AGN at $z\sim 1$. In principle, the evolution curves shown in
Fig.~\ref{broz} can also be used to predict which $r_0$ value a given
XMM-COSMOS subsample should have as a function of redshift, allowing
a proper comparison between measurements obtained at different
redshifts. Indeed, the correlation length of AGN in the $z>1$ sample
(median $z\sim1.5$) is expected to evolve to $r_0\sim7.5\;h^{-1}$ Mpc
by $z\sim0.7$, i.e. the median redshift of the $z<1$ sample, whose
correlation length has been measured as varying between 8.0 and 5.2
$h^{-1}$ Mpc, depending on the inclusion of the $z=0.36$
structure. Given this uncertainty and the large errorbars in
Fig.~\ref{broz}, it is still difficult to claim that objects at
redshift greater or smaller than 1 are sampling different environments.

\subsection{Estimating the AGN lifetime}\label{life}

Under simple assumptions it is possible to put limits on the AGN
lifetime at any given redshift. Following Martini \& Weinberg
(\cite{mw01}), we assumed that the AGN in our sample reside within
halos above a given mass threshold and that each halo hosts at most
one active AGN at a time. The AGN lifetime $t_Q$ can then be estimated
with the following relation:

\begin{equation}
\Phi(z) = \int_{M_{min}}^{\infty}dM\frac{t_Q}{t_H(M,z)}n(M,z),
\label{tq}
\end{equation}
where $\Phi(z)$ is the comoving space density of AGN above a given luminosity,
$M_{min}$ the minimum mass of the halos hosting an AGN, $n(M,z)$ the 
comoving space density of halos of mass $M$ at redshift $z$,
and $t_H(M,z)$ is the lifetime of halos of mass $M$ at redshift $z$.

The definition of halo lifetime is somewhat ambiguous since halos are
continuously accreting matter. Martini \& Weinberg (\cite{mw01})
defined $t_H(M,z)$ as the median time interval for a halo of mass M to
be incorporated into a halo of mass 2M and used the extended
Press-Schechter formalism to calculate it. To a first approximation
$t_H(M,z)\sim t_U(z)$, where $ t_U(z)$ is the Hubble time at redshift
$z$. With these approximations Eq.~\ref{tq} can be rewritten as

\begin{equation}
t_Q(z)=t_u(z)\frac{\Phi(z)}{\Phi_H(z)},
\label{tq2}
\end{equation}
where $\Phi_H(z)=\int_{M_{min}}^{\infty}dM\;n(M,z)$ is the comoving
space density of halo with mass above $M_{min}$.

Since $M_{min}$ is known from the comparison between the halo and the
AGN correlation length, it is straightforward to estimate $\Phi_H(z)$
from the number of halos with $M>M_{min}$ within the Millennium
simulation box. For halos with $M>2.5 \times 10^{12} \; h^{-1} \;
M_{\odot}$, where XMM-COSMOS AGN at $z\sim 1$ reside, the space
density is $\Phi_H(z)= 10^{-3}\;h^3$ Mpc$^{-3}$ (see
Fig.~\ref{nr0}). For the cosmology adopted here the Hubble time at
$z=1$ is $\sim 6.3$ Gyr.

The comoving space density of XMM-COSMOS AGN has been estimated by
considering literature X-ray luminosity function of AGN selected in
the 2-10 keV band, which should therefore include unobscured, as well
as moderately obscured, objects as the objects populating our
sample. Once accounting for band effects\footnote{an X-ray photon
index of $\Gamma=1.9$ is assumed}, the median luminosity of our
$z\sim 1$ sample (logL=43.7 in the 0.5-10 keV band; see Table~1)
translates into a 2-10 keV luminosity of logL$\sim 43.5$. At these
luminosities, $z\sim 1$ AGN in the La Franca et al. (2005) XLF have a
space density of $\sim 3\times 10^{-4}\;h^3$ Mpc$^{-3}$. A similar
value for the AGN density is obtained when using the XLF by Ueda et
al. (2003).
Therefore, by considering an AGN density of $\sim 1.8\times 10^{-4}\;h^3$
Mpc$^{-3}$ (obtained by rescaling the La Franca et al. space density
by the fraction of objects with $I_{AB}<23$, as is the case for our
selection), a duty cycle $t_Q/t_H$ of 0.18 is obtained, which
translates into an AGN lifetime of $\sim 1.1$ Gyr.

This estimated lifetime is more than one order of magnitude longer than
that estimated by Porciani et al. (2004) for bright optical QSOs at
$z\sim 1$ in the 2QZ survey. The difference in the measured lifetime
is essentially due to the difference between the space density of
XMM-COSMOS AGN and 2QZ QSOs at $z\sim 1 \;(\approx 2\times 10^{-4}\;
\rm{vs}\;\approx 10^{-5}\;h^3$ Mpc$^{-3}$, see Table~1 in Porciani et
al. 2004). Such a difference is, on the other hand, expected given the
relatively bright limiting magnitude ($m_B\sim 20.8$) of the 2QZ
sample that, in addition, does not include obscured AGN. The
estimated lifetime for XMM-COSMOS AGN is significantly shorter than
the $\sim 8$ Gyr time span between $z=1$ and $z=0$. This, in
combination with the estimate that XMM-COSMOS AGN will cluster with
$r_0=8$ at $z=0$, depicts a consistent scenario in which XMM-COSMOS AGN
will switch off by $z=0$, leaving relic (dormant) supermassive black
holes in local elliptical galaxies. 


\section{Conclusions}

We have studied the clustering properties of 538 moderately luminous
AGN at $z=0.2-3$ in the 2 deg$^2$ COSMOS field, selected in the X-rays
and spectroscopically identified to $I_{AB}<23$. Our main results can
be summarized as follows:\\

1. The projected correlation function $w(r_p)$ on scales
$r_p=0.3-40\;h^{-1}$ Mpc can be approximated by a power law with
correlation length $r_0=8.6\pm0.5 \;h^{-1}$ and slope
$\gamma=1.9\pm0.1$ (Poisson errors; bootstrap errors are a factor of
$\sim 2$ larger). This represents the most significant measurement of
clustering of X-ray selected AGN to date.\\

2. Part of the signal, in particular an excess on projected scales
$r_p=5-15\;h^{-1}$ Mpc, is due to a large scale structure at
$z=0.36$. When excluding this structure or computing $w(r_p)$ for
objects in a narrower redshift interval around $z\sim 1$, the
correlation length decreases to $r_0=5-6\;h^{-1}$ Mpc, similar to what
is observed in large samples of optically selected QSOs at the same
redshift.\\

3. Objects with different absorption properties do not show
significant evidence for different clustering properties. Broad line
AGN are consistent with inhabiting the same environments of non-broad
line AGN. Similar results are obtained when considering X-ray absorbed
and X-ray unabsorbed AGN.\\

4. No significant difference is found in the clustering properties of
objects at redshifts below or above 1.\\

5. The correlation length measured for XMM-COSMOS AGN at $z\sim 1$ is
similar to that of early type galaxies and luminous infrared galaxies
at the same redshift. This, in agreement with other studies, suggests
that $z\sim 1$ moderately luminous AGN are found preferentially in
massive ($M\gtrsim3\times10^{10}\;M_{\odot}$) galaxies.\\

6. By using public halo catalogs from the Millennium simulation, we
estimated XMM-COSMOS AGN to reside within dark matter halos of mass
$M\gtrsim2.5\times10^{12}\;h^{-1}\;M_{\odot}$.\\

7. According to a simple {\it conserving scenario} for clustering
evolution, the relics of $z\sim 1$ AGN are expected to be hosted by
local bright $L\sim L_\star$ ellipticals by z=0.\\

8. By combining the number density of XMM-COSMOS AGN with that of the
hosting dark matter halos, we estimated an AGN duty cycle of 0.1, which
translates into an AGN lifetime of $\sim 1$ Gyr. The estimated
lifetime is more than one order of magnitude longer than
estimated for optically bright QSOs at the same redshift. This is
mainly due to the higher number density of AGN found in X-ray selected
samples.\\

\acknowledgements

This work is based on observations obtained with XMM-Newton, an ESA
Science Mission with instruments and contributions directly funded by
ESA Member States and the USA (NASA). We gratefully acknowledge the
contribution of the entire COSMOS collaboration ({\tt
http://www.astro.caltech.edu/\~cosmos}). In Italy, the XMM-COSMOS
projects is supported by ASI-INAF and PRIN/MIUR under grants
I/023/05/00 and 2006-02-5203. The zCOSMOS ESO Large Program Number
175.A-0839 is acknowledged. RG thanks Carlo Nipoti, Federico Marulli,
Enzo Branchini, and Lauro Moscardini for stimulating discussions. The
referee is acknowledged for providing useful comments. The Millennium
Simulation databases used in this paper and the web application
providing online access to them were constructed as part of the
activities of the German Astrophysical Virtual Observatory.


\end{document}